\pdfoutput=1
\documentclass[reprint,aps,fleqn,usenatbib,nofootinbib]{revtex4-1}
\usepackage{graphicx}
\usepackage{amsmath,amssymb}
\usepackage{lastpage}

\usepackage{hyperref}
\usepackage[all]{hypcap}
\usepackage{float}
\usepackage{subfigure}
\usepackage{afterpage}
\usepackage[usenames]{color}
\usepackage{yfonts}
\usepackage{mathrsfs}
\usepackage{upgreek}
\usepackage{gensymb}

\begin{document}

\preprint{APS/123-QED}

\title{Fifth force constraints from the separation of galaxy mass components}

\author{Harry Desmond}%
 \affiliation{Astrophysics, University of Oxford, Denys Wilkinson Building, Keble Road, Oxford OX1 3RH, United Kingdom}
 \email{harry.desmond@physics.ox.ac.uk}
\author{Pedro~G.~Ferreira}%
 \affiliation{Astrophysics, University of Oxford, Denys Wilkinson Building, Keble Road, Oxford OX1 3RH, United Kingdom}
\author{Guilhem Lavaux}%
 \affiliation{Sorbonne Universit{\'é}, CNRS, UMR 7095, Institut d'Astrophysique de Paris, 98 bis bd Arago, 75014 Paris, France}
 \affiliation{Sorbonne Universit{\'é}s, Institut Lagrange de Paris (ILP), 98 bis bd Arago, 75014 Paris, France}
\author{Jens Jasche}%
 \affiliation{The Oskar Klein Centre, Department of Physics, Stockholm University, Albanova University Center, SE 106 91 Stockholm, Sweden}
 \affiliation{Excellence Cluster Universe, Technische Universit{\"ä}t M{\"ü}nchen, Boltzmannstrasse 2, D-85748 Garching, Germany}

\date{\today}


\begin{abstract}
One of the most common consequences of extensions to the standard models of particle physics or cosmology is the emergence of a fifth force. While generic fifth forces are tightly constrained at Solar System scales and below, they may escape detection by means of a screening mechanism which effectively removes them in dense environments. We constrain the strength $\Delta G/G_N$ and range $\lambda_C$ of a fifth force with Yukawa coupling arising from a chameleon- or symmetron-screened scalar field -- as well as an unscreened fifth force with differential coupling to galactic mass components -- by searching for the displacement it predicts between galaxies' stellar and gas mass centroids. Taking data from the \textit{Alfalfa} survey of neutral atomic hydrogen (HI), identifying galaxies' gravitational environments with the maps of~\cite{Desmond} and forward-modelling with a Bayesian likelihood framework, we find, with screening included, $6.6\sigma$ evidence for $\Delta G>0$ at $\lambda_C \simeq 2$ Mpc. The maximum-likelihood $\Delta G/G_N$ is $0.025$. A similar fifth-force model without screening gives no increase in likelihood over the case $\Delta G = 0$ for any $\lambda_C$. Although we validate this result by several methods, we do not claim screened modified gravity to provide the only possible explanation for the data: this conclusion would require knowing that the signal could not be produced by ``galaxy formation'' physics. We show also the results of a more conservative -- though less well motivated -- noise model which yields only upper limits on $\Delta G/G_N$, ranging from $\sim10^{-1}$ for $\lambda_C \simeq 0.5$ Mpc to $\sim \: \text{few} \times 10^{-4}$ at $\lambda_C \simeq 50$ Mpc. Corresponding models without screening receive the somewhat stronger bounds $\sim \: \text{few} \times 10^{-3}$ and $\sim \: \text{few} \times 10^{-4}$ respectively. We show how these constraints may be improved by future galaxy surveys and identify the key features of an observational programme for directly constraining fifth forces on scales beyond the Solar System. This paper provides a complete description of the analysis summarised in \cite{Desmond_PRL}.
\end{abstract}

\maketitle

\section{Introduction}
\label{sec:intro}

While $\Lambda$CDM has so far passed most of its tests with flying colours, the possibility remains that it requires modification either at levels of sensitivity currently unattainable, or in regions of parameter space so far unexplored. In the past few decades, this has motivated a burgeoning programme aimed at formulating theoretical alternatives to concordance cosmology, developing phenomenological frameworks to explore their consequences and devising novel experimental and observational probes.

A generic feature of fundamental extensions to the standard model is the introduction of new dynamical degrees of freedom. These may be produced by substituting dimensional parameters by dynamical fields (e.g. masses \cite{Weinberg}, dark energy \cite{Ratra} or the gravitational constant \cite{Brans,Wetterich}), and also describe the effects of higher derivatives in the Lagrangian as well as extra dimensions. Any extension to the Einstein-Hilbert action will necessarily involve new fields~\citep{Clifton}, and high-energy completions of known physics produce a plethora of new degrees of freedom which may operate on a wide range of scales. Given the ubiquity of novel fields in both gravitational and particle-based extensions to the standard model, seeking new physics is largely tantamount to devising methods to detect additional scalars, vectors or tensors.

Any new field couples naturally to the Ricci scalar $R$ in the gravitational action, and hence contributes both to the Universe's energy-momentum and to the expression for the force. The simplest and most common case is a single scalar field with standard kinetic energy, potential $V(\phi)$, and coupling $\phi^2R$. This modifies the potential generated by a point mass $M$ to
\begin{equation}
\Phi_\text{tot} = -\frac{G_NM}{r}\left(1+\frac{\Delta G}{G_N}e^{-mr}\right) = \Phi_N - \frac{\Delta G M}{r}e^{-mr},  \label{eq:ff}
\end{equation}
where $G_N$ is the bare (Newtonian) gravitational constant, $\Phi_N$ is the Newtonian potential, and $m$ and $\Delta G$ are new parameters describing the mass of the field and the strength of the new interaction respectively. In theoretical terms, $m$ is set by the potential as $m \sim d^2V/d\phi^2$ while $\Delta G$ depends on the strength of the non-minimal coupling and the background value of the field relative to the Planck mass. The bare mass may be altered to the effective mass $m_\text{eff}$ given screening (see below), which sets the range of the field as $\lambda_C \simeq 1/m_\text{eff}$. Eq.~\ref{eq:ff} describes a violation of the inverse-square law by the new field and the introduction of a \emph{fifth force} ($F_5$) which adds to the standard gravitational force with a Yukawa form. General Relativity (GR) is recovered in the limit $\Delta G\rightarrow 0$, and also in the case where the field's mass is sufficiently large for the fifth force to be strongly Yukawa-suppressed ($m \rightarrow \infty$, $\lambda_C \rightarrow 0$). In the case where the field is light ($m \rightarrow 0$, $\lambda_C \rightarrow \infty$), the sole effect of the new field is to renormalise the gravitational constant to $G \equiv G_N + \Delta G$. Note that the existence of fifth forces requires only the presence of scalar fields, and hence does not necessarily require modifications to the standard model: the Higgs field for example leads a fifth force with sub-nuclear range.

The empirical programme of probing fifth forces (or, relatedly, violations of the inverse-square law) has been ongoing for around two decades; see~\cite{Adelberger} for a review. At small scales ($\sim 10^{-5}-10^{-2}$ m), the strongest limits are set by the E\"{o}t-Wash experiment which involves precise measurement of the force between two nearby objects~\citep{Adelberger_0, Hoyle}, supplemented by high frequency torsion oscillators~\citep{Colarado} and microcantilevers~\citep{Stanford} at even shorter distances. These constraints are $\Delta G/G_N \lesssim 10^{-6}$ at $\lambda_C = 10^{-5}$ m and $\Delta G/G_N \lesssim 10^{-2}$ at $\lambda_C = 10^{-2}$ m. At astrophysical scales, constraints derive from the Earth-moon distance probed by lunar laser ranging ($\Delta G/G_N \lesssim 10^{-10}$ at $\lambda_C = 10^9$ m) and planetary orbits ($\Delta G/G_N \lesssim \text{few} \times 10^{-9}$ at $\lambda_C = 10^{11}$ m)~\citep{Williams, Samain}. No direct constraint exists at larger scales due to the difficulty of predicting precisely the motions of masses beyond the Solar System, even under the assumption of standard gravity.

Unlike an equivalence principle test, a fifth-force search is restricted to ranges $\lambda_C$ comparable to the size of the system under investigation. Ranges much smaller than the system will go unnoticed as the scalar field will not couple its various parts; conversely, if the range is much larger, the entire system simply feels a renormalised Newton's constant $G$ and the inverse-square law is satisfied. This renders tests which rely on the distance-dependence of the force between masses, such as E\"{o}t-Wash and lunar laser ranging, inoperable. However, because a light scalar field affects timelike but not null geodesics it can be probed by means of the Eddington light-bending parameter $\gamma = (1-\Delta G/G_N)/(1+\Delta G/G_N)$ in the Parametrised Post-Newtonian framework. $\gamma$ has been constrained to $\lesssim 10^{-5}$ by the Cassini satellite~\citep{Bertotti}, requiring $\Delta G/G_N \lesssim 10^{-5}$. This by itself is sufficient to force a putative scalar into astrophysically and cosmologically uninteresting regions of parameter space, as its contribution to the Klein-Gordon equation cannot be more than a fraction $\mathcal{O}(10^{-3})$ that of standard gravity.

Nevertheless, there exist ways in which a fifth force may escape these constraints and still have an impact at larger scales. In a number of theories (for example generalised scalar-tensor theories and massive gravity) the properties of $F_5$ depend on the local environment: in high density regions such as the Milky Way its strength or range becomes small, while it remains operative in the lower density environments of dwarf galaxies and the Universe as a whole. This phenomenon is called ``screening'' (see~\cite{Jain_Khoury,Khoury_review} for reviews), and comes in three basic flavours. The \emph{chameleon} mechanism~\citep{chameleon} arises when the effective mass $m_\text{eff}$ becomes dependent on the local density (and thus on $\nabla^2\Phi$), such that in denser regions $m_\text{eff}$ becomes large and the fifth force short range, while in less dense regions (or on cosmological scales) $m_\text{eff}\rightarrow0$ and the fifth force reappears. In \emph{kinetic} screening (e.g. K-mouflage;~\cite{kinetic}), the scalar interaction is governed by a kinetic function $\sim(\partial \phi)^2$, so that $\phi$ becomes nonlinear, and hence effectively decouples, when its gradient is large. Finally, in the {\it Vainshtein}~\citep{Vainshtein} and {\it symmetron}~\citep{Hinterbichler} mechanisms, the coupling of the field to matter depends on the local environment, such that $\Delta G\rightarrow 0$ near massive bodies. In each case, Solar System tests would be expected to probe the screened regime and hence yield the GR result.

In the presence of screening, new scalar fields are possible with Compton wavelengths ranging from microscopic to cosmological values. This renders observationally viable a broad class of gravitational actions, including those that may be expected to form the effective low-energy limits of UV-complete theories. Further, screening holds opens the possibility for cosmic acceleration to receive a dynamical explanation, potentially circumventing the cosmological constant problem~\citep{Carroll}, and sanctions the search for fifth-force phenomenology on all scales. Crucially, however, tests of screening require unscreened objects, which are the only ones to manifest $F_5$.

The aim of this paper is to constrain fifth forces in galaxies. Not only will this provide direct information on scales much larger than the Solar System ($\sim0.4-50$ Mpc), it will also enable constraints in the presence of screening. This is possible due to the wide range of densities and environments of galaxies in the local Universe, including those with masses much lower than the Milky Way and in environments much sparser, which would remain unscreened even for small background values of a novel scalar field.

We restrict ourselves here to chameleon and symmetron screening,\footnote{Similar behaviour is expected for the environmentally-dependent dilaton~\cite{Brax}.} which are conceptually simpler and more amenable to testing through the signal we wish to explore. In these mechanisms, whether or not an object is screened depends on the value of its Newtonian potential $\Phi$ \cite{Brax_2}. For a single spherical source this is deducible analytically as a consequence of the fact that chameleon and symmetron charge is contained in a thin shell near the object's surface (the ``thin-shell effect'';~\cite{chameleon, Khoury_LesHouches}). For multiple sources (and taking into account the possibility of environmental as well as self-screening), numerical simulations show that $\Phi = \Phi_\text{in} + \Phi_\text{ex}$ -- where $\Phi_\text{in}$ is the Newtonian potential generated by the object itself at its surface, and $\Phi_\text{ex}$ that due to its surroundings -- remains the best proxy for the degree of screening, as measured by differences between dynamical mass, which is subject to the fifth force, and lensing mass which is not~\citep{Zhao_1, Zhao_2, Cabre, Falck}. In addition to fifth-force strength $\Delta G$ and range $\lambda_C$, then, generic chameleon and symmetron models come also with a critical value of $\Phi$, $\Phi_c$, at which screening kicks in. To good approximation, objects with $|\Phi|>|\Phi_c|$ are in sufficiently dense environments to be screened ($m_\text{eff} \rightarrow \infty$), while those with $|\Phi|<|\Phi_c|$ are not ($m_\text{eff} \rightarrow 0$).\footnote{An alternative screening criterion (e.g.~\cite{Cabre}) is $|\Phi_\text{in}| > |\Phi_c|$ \emph{or} $|\Phi_\text{ex}| > |\Phi_c|$. As this is more difficult to satisfy than $|\Phi_\text{in}| + |\Phi_\text{ex}| > |\Phi_c|$, our screened fractions may be considered upper bounds, making our constraints (which benefit from more \emph{un}screened galaxies in which the predicted signal is $>$0) conservative in this regard.} For viable and interesting theory parameters, $|\Phi_c|$ takes values in the range $\sim 10^{-5}-10^{-8}$ ($c \equiv 1$), putting some galaxies in the screened and others the unscreened regime.

Thin-shell screening possesses a range of intra-galaxy signals, all of which derive from a common basic effect. Within an unscreened galaxy, gas and dark matter interact via a fifth force with neighbouring unscreened mass, which leads to an effective increase in Newton's constant $\Delta G = 2\beta^2G$ if the scalar is light and coupled to matter by $\nabla^2\phi = 8\pi\beta G\rho$. (If the galaxy is marginally screened there is an addition thin-shell factor, but this is not important for the cases of practical interest here.) Main sequence stars, however, are themselves massive objects with surface Newtonian potentials $|\Phi_\text{in}|\sim10^{-6}$, and hence for $|\Phi_c| \lesssim 10^{-6}$ they self-screen and do not feel $F_5$. The effective equivalence principle violation \cite{Hui} caused by the differential sensitivity of stars, gas and dark matter to the fifth force has four observable consequences \cite{Jain_Vanderplas}: 1) A displacement between the stellar and gas disk in the direction of $F_5$, 2) warping of the stellar disk, 3) enhancement of the rotation velocity of the gas relative to the stars, and 4) asymmetry in the stellar disk's rotation curve. We focus in this paper on the first. By collating neutral atomic hydrogen (HI) and stellar (optical) data for a large sample of galaxies, calculating their gravitational environments and forward-modelling the expected signal for given theory parameters $\{\Delta G, \lambda_C, \Phi_c\}$, we statistically constrain a fifth force screened by any thin-shell mechanism. We also provide constraints in the case without screening but where the fifth force is simply coupled differently to stars, gas and dark matter.

The structure of this paper is as follows. In Sec.~\ref{sec:derivation} we derive the form of the HI-optical offset for a given coupling of $F_5$ to stars and gas, and in Sec.~\ref{sec:data} we describe the survey data that we use to search for it. Sec.~\ref{sec:method} details our method: first our mapping of the potential and fifth-force acceleration fields across the local Universe, second the calculation from these of the signal expected for given $\{\Delta G, \lambda_C, \Phi_c\}$, and third the likelihood framework we use to constrain the theory parameters by Markov Chain Monte Carlo (MCMC). Sec.~\ref{sec:results} presents our results and forecasts improvements from future data. In Sec.~\ref{sec:discussion} we provide further information on our noise model, document the systematic uncertainties in our analysis, compare our work to the literature and offer suggestions for future study. Finally, Sec.~\ref{sec:conc} concludes.

\section{The signal: gas--star offsets in the presence of a fifth force}
\label{sec:derivation}

Consider an unscreened galaxy in a chameleon- or symmetron-screened theory subject to an external Newtonian acceleration $\vec{a}$ and a Newtonian acceleration $\vec{a}_5$ due to unscreened matter within $\lambda_C$. Take $|\Phi_c| \lesssim 10^{-6}$ so that stars self-screen. Under the step-function approximation of Eq.~\ref{eq:ff}, unscreened objects with separation $r < \lambda_C$ interact through standard gravity and $F_5$, while those with $r > \lambda_C$ interact only through the former. Thus the unscreened gas and dark matter will effectively couple to the mass generating $\vec{a}_5$ with enhanced gravitation constant $G_N + \Delta G$, and hence feel a total acceleration
\begin{equation} \label{eq:agas}
\vec{a}_{g,DM} = \vec{a} + \frac{\Delta G}{G_N} \: \vec{a}_5.
\end{equation}

In order for the stars to remain associated with the gas and dark matter, their insensitivity to $\vec{a}_5$ must be compensated by an acceleration due to a separation between their centre of mass and that of the dark matter halo. If the stellar centroid lags behind the halo centre in the direction of $\vec{a}_5$ by a distance $r_*$, its total acceleration will be
\begin{equation} \label{eq:astar}
\vec{a}_* = \vec{a} + \frac{G_N M(<r_*)}{r_*^2} \: \hat{r}_*,
\end{equation}
where $M(<r)$ is the dark matter plus gas mass enclosed by a sphere of radius $r$ around the halo centre. Requiring $\vec{a}_{g,DM} = \vec{a}_*$ so that all parts of the galaxy move together, and distinguishing explicitly between the cases where the galaxy itself is screened or unscreened, we find that the equilibrium offset $\vec{r}_*$ satisfies
\begin{align}\label{eq:rstar}
&\frac{M(<r_*)}{r_*^2} \: \hat{r}_* = \vec{a}_5 \: \frac{\Delta G}{G_N^2}, &\quad |\Phi| < |\Phi_c|\\ \nonumber
&\vec{r}_* = 0, &\quad |\Phi| > |\Phi_c|&
\end{align}
where $\Phi$ is the potential of the galaxy as a whole. This is the equation we will solve in Sec.~\ref{sec:method}, as a function of $\Delta G$, $\lambda_C$ and $\Phi_c$, to calculate the expected signal for the observed galaxies.

Although we have couched our discussion in terms of thin-shell screening, the phenomenon we describe would also result from a generic fifth force, screened or unscreened, that couples differently to different galaxy mass components. In that case, $\Delta G$ in Eq.~\ref{eq:rstar} would be related to that in Eq.~\ref{eq:ff} by a factor of the normalised coupling coefficient of $F_5$ to gas and dark matter minus that to stars.

\section{Observational data}
\label{sec:data}

Our observational catalogue of HI-optical displacements is derived from the \textit{Alfalfa} survey~\citep{Giovanelli_1, Giovanelli_2, Saintonge, Kent, Martin, Haynes}, a second-generation blind HI survey of $\sim 30,000$ extragalactic sources conducted with the Arecibo Observatory. \textit{Alfalfa} is sensitive to objects of mass $10^6 \lesssim M_\text{HI}/M_\odot \lesssim 10^{10.8}$, and covers over $7000 \: \text{deg}^2$ out to $z \simeq 0.06$. More than 98\% of HI detections were associated with optical counterparts (OCs) by interactively examining DSS2 and SDSS images, in addition to objects in the Nasa Extragalactic Database (NED).

We cut the sources designated as ``priors'' (quality flag 2), which have signal to noise ratio $s\lesssim6.5$ but are nevertheless matched to OCs with optical redshifts consistent with those derived from HI, as well as those estimated to be high velocity clouds in the Milky Way (quality flag 9). We also remove galaxies with an offset angle between HI and OC centroids of $>2$ arcmin, which is likely caused by poor signal to noise. These are similar cuts to those employed in previous work along these lines~\cite{Vikram}. Objects at larger distance $D$ have larger physical uncertainty in HI-optical displacement for given angular uncertainty $\theta$, and we will find in Sec.~\ref{sec:results} that for realistic $\theta$ values (see Sec.~\ref{sec:uncertainties}) objects with $D>100$ Mpc do not provide further constraining power on $\{\Delta G, \lambda_C, \Phi_c\}$. We therefore cut the catalogue at 100 Mpc, yielding 10,822 galaxies in the final sample.

\section{Method}
\label{sec:method}

We construct a Bayesian framework for constraining $\{\Delta G, \lambda_C, \Phi_c\}$ by comparing forward-modelled $\vec{r}_*$ values to those observed in \textit{Alfalfa} galaxies. To do so, we derive or specify probability distributions for the input parameters necessary to solve Eq~\ref{eq:rstar}, and use Monte Carlo sampling to propagate their uncertainties into the predicted $\vec{r}_*$. We thus create a likelihood function for that signal, and score models by applying Bayes' theorem and an MCMC algorithm to it.

To simplify our inference, we first approximate the exponential in Eq.~\ref{eq:ff} by a step function at $\lambda_C$,\footnote{As our gravitational maps become inaccurate beyond $\sim200$ Mpc, this approximation is necessary for larger fifth-force ranges because we cannot include contributions from sources $\gtrsim2\lambda_C$ from test points $\sim100$ Mpc from us. With future maps extending to higher redshift it will be preferable to treat Eq.~\ref{eq:ff} more exactly.} and second impose the relation between $\lambda_C$ and screening threshold $\Phi_c$ implied by a specific chameleon-screened theory, the Hu-Sawicki~\citep{Hu_Sawicki} model of $f(R)$ gravity~\citep{fR, fR2}. In this case, both $\lambda_C$ and $\Phi_c$ are set by the background value of the scalar field, which is equal to the derivative of the function $f$ at the cosmological value $R_0$ of the Ricci scalar $R$, $f_{R0} \equiv df/dR|_{R_0}$ (also the background scalar field value $\phi_0$):
\begin{subequations}
\label{eq:f(R)}
\begin{align}
&\lambda_C = 32 \: \sqrt{f_{R0}/10^{-4}} \: \text{Mpc}, \label{eq:f(R)1}\\
&|\Phi_c| = \frac{3}{2} f_{R0}, \label{eq:f(R)2}\\ \nonumber
&\text{so that}\\
&|\Phi_c| = \frac{3}{2} \times 10^{-4} \left(\frac{\lambda_C}{32 \: \text{Mpc}}\right)^2. \label{eq:f(R)3}
\end{align}
\end{subequations}
Qualitatively, these relations are also of more general applicability with $\lambda_C$ interpreted in terms of the self-screening parameter $\phi_0/(2\beta M_\text{pl})$, often called $\chi_c$. Although in the most general screening theory $\lambda_C$ and $\Phi_c$ are likely unrelated, we will find the observational data to possess insufficient information for a three dimensional inference, and hence impose Eq.~\ref{eq:f(R)} by fiat. We leave it to future modelling with better data to open the parameter space further. The model that we seek to test, then, is fully specified by $\Delta G$ and $\lambda_C$.

For given $\{\Delta G, \lambda_C\}$, we derive the gravitational inputs to the predicted $\vec{r}_*$ ($\Phi$ and $\vec{a}_5$) in Secs~\ref{sec:maps} and~\ref{sec:a5} and the galactic input ($M(<r)$) in Sec.~\ref{sec:signal}. Sec.~\ref{sec:MC} details our sampling method, Sec.~\ref{sec:likelihood} describes our likelihood function, and Sec.~\ref{sec:convergence} presents convergence and consistency tests.

\subsection{Mapping $\Phi$ and $\vec{a}$}
\label{sec:maps}

We determine an object's degree of screening by means of its Newtonian potential $\Phi$. This is the sum of an internal component $\Phi_\text{in}$, due to the object's own mass, and an external component $\Phi_\text{ex}$ due to its environment. $\Phi_\text{in}$ can be estimated either from the object's mass and size, or from its characteristic velocity. The simplest estimate\footnote{This is how $\Phi_\text{in}$ was calculated in the simulations which established $\Phi_\text{in} + \Phi_\text{ex}$ to be a useful proxy for chameleon screening~\citep{Zhao_1, Zhao_2, Cabre}. By Eq.~\ref{eq:Vsum}, however, the Newtonian potential near the centres of halos where galaxies live may be smaller by an order of magnitude or more. Switching to this would reduce screened fractions and boost the predicted signal; using Eq.~\ref{eq:Vin} is therefore conservative. Nevertheless, given more information on the internal dynamics of test galaxies it would be preferable to develop a more precise estimator for the degree of self-screening.} is
\begin{equation}
|\Phi_\text{in}| = \frac{G M_\text{vir}}{R_\text{vir}} = V_\text{vir}^2. \label{eq:Vin}
\end{equation}
$\Phi_\text{ex}$ on the other hand -- along with the acceleration $\vec{a}$, which has only an external component\footnote{We assume that galaxies are small and uniform enough, relative to the fifth-force range, for tidal fields within them to be unimportant. This may incur error for $\lambda_C$ at the small end of our range of interest (400 kpc): at this point the acceleration of one mass component within a galaxy may receive a relatively significant contribution from others, making $\vec{a}_5$ dependent on the structure of the galaxy itself. We will see in any case that the $\Delta G$ constraints for $\lambda_C \sim 400$ kpc are relatively weak.} -- is a non-local function of the full density field in space. The remainder of this subsection is devoted to calculating it.

Our starting point is the method of~\cite{Desmond} which sums the contributions from three sources: mass associated with survey galaxies, halos hosting galaxies above the survey magnitude limit, and matter situated outside collapsed structures. Briefly, the procedure for determining these three parts is as follows (further details in that paper):

\begin{enumerate}

\item{} Estimate the total dynamical mass distributions of galaxies in 2M++~\cite{2M++}, an all-sky redshift survey complete to $K=12.5$ out to $z \simeq 0.05$. This is achieved by associating an NFW halo to each 2M++ galaxy using the technique of inverse abundance matching (AM;~\cite{Kravtsov,Conroy,Reddick}) -- which assumes a nearly monotonic relation between galaxy luminosity (in this case $K$-band) and a function of halo mass $M_\text{vir}$ and concentration $c$, known as the `proxy' -- to assign galaxy luminosities to halos in an N-body simulation. We adopt the particular AM model of~\cite{Lehmann}, which uses the proxy $V_\text{vir} \: (V_\text{max}/V_\text{vir})^\alpha$, with best-fit value $\alpha = 0.6$, and a Gaussian uncertainty $\sigma_\text{AM}$ with best-fit value 0.16 dex. This reproduces the observed clustering of galaxies from the simulated clustering of halos (in our case \textsc{rockstar} \cite{Rockstar} halos in the \textsc{darksky-400} \cite{DarkSky} box). The first part of $\Phi_\text{ex}$ and $\vec{a}$, denoted by subscript `vis', are then as follows:
\begin{equation}
\Phi_\text{vis} = -\sum_i \: \frac{G M_\text{vir,i}}{r_i} \: \frac{\ln(1+\frac{c_i r_i}{R_\text{vir,i}})}{\ln(1+c_i)-\frac{c_i}{1+c_i}}, \label{eq:Vsum}
\end{equation}
\begin{equation}
\vec{a}_\text{vis} = -\sum_i \: \frac{G M_\text{vir,i}}{r_i} \: \frac{\frac{1}{r_i} \ln(1+ \frac{c_i r_i}{R_\text{vir,i}}) - \frac{c_i/R_\text{vir,i}}{1+c_i r_i/R_\text{vir,i}}}{\ln(1+c_i)-\frac{c_i}{1+c_i}} \: \hat{r}_i, \label{eq:asum}
\end{equation}
for source halo $i$ at distance $r_i$ from the test point.

\item{} Correct $\Phi_\text{vis}$ and $\vec{a}_\text{vis}$ for halos hosting galaxies below the 2M++ detection threshold. We supply $K$-band luminosities for \textsc{darksky-400} halos by AM, and calculate their apparent magnitude to partition them into subsets visible and invisible to the 2M++ survey. We calculate the ``correction factors'' $\vec{c} \equiv \{c_\Phi, c_a, c_\theta\}$, at the locations of \textsc{darksky-400} objects, required to convert $\Phi_\text{vis}$ and $\vec{a}_\text{vis}$ to $\Phi_\text{hal}$ and $\vec{a}_\text{hal}$, sourced by all halos with $M_\text{vir} > 7.63 \times 10^{10} \: M_\odot$ (the 1000-particle resolution limit of the simulation;~\cite{Diemer_Kravtsov}):
\begin{equation}
\Phi_\text{hal} \equiv c_\Phi \Phi_\text{vis}; \; \; |\vec{a}_\text{hal}| \equiv c_a |\vec{a}_\text{vis}|; \; \; c_\theta \equiv \frac{\vec{a}_\text{hal} \cdot \vec{a}_\text{vis}}{|\vec{a}_\text{hal}| |\vec{a}_\text{vis}|}.
\end{equation}
To apply this correction in the real Universe, we correlate these factors in the simulation with the observational proxies $d$ (distance to the object), $N_{10}$ (number of objects visible to 2M++ within 10 Mpc) and $\Phi_\text{vis,10}$ (potential due to objects visible to 2M++ within 10 Mpc). We calculate $\{d, N_{10}, \Phi_\text{vis,10}\}$ for each real test galaxy and assign it the $\vec{c}$ values of a randomly-chosen \textsc{darksky-400} halo in the same bin of proxy values.

\item{} Add the contributions $\Phi_\text{sm}$ and $\vec{a}_\text{sm}$ from the remaining matter, not associated with resolved halos, by estimating the mass in linear and quasi-linear modes of the density field (subscript `sm' denotes `smooth'). This is achieved by means of a Bayesian reconstruction of the $z \lesssim 0.05$ density field with the Bayesian Origin Reconstruction from Galaxies (BORG) algorithm~\citep{Jasche_10, Jasche_15, Jasche_Wandelt_12, Jasche_Wandelt_13, Lavaux}, which propagates information on the number densities and peculiar velocities of 2M++ galaxies assuming $\Lambda$CDM cosmology (with best-fit Planck parameter values) and a bias model. The parameters of the bias model are not known a priori but fitted directly to the data. We use the brand-new particle mesh BORG algorithm~\cite{BORG_PM}, which improves in accuracy on the previous 2-loop perturbation theory determination used in~\cite{Desmond}.

For our purposes, the output of this algorithm is a posterior distribution of density fields for the local $(677.7 \: h^{-1} \: \text{Mpc})^3$ volume, with a grid scale of $\Delta r = 2.65 \: h^{-1}$ Mpc. We approximate each field as a set of point masses at the grid cell centres. Although sufficient for $\lambda_C \gtrsim 10$ Mpc, when several grid cells are contained within $\lambda_C$ around each test point, this discretisation will generate excessive shot noise at smaller $\lambda_C$ due to the relative sparsity of source masses. Test points equidistant from neighbouring source masses may enclose none within $\lambda_C$, and hence artificially receive $\Phi_\text{ex} = \vec{a} = r_* = 0$. To rectify this, we define a minimum value for the radius out to which source objects are considered, $R_\text{max}$, of $R_\text{thresh} =10\;\text{Mpc}\approx 2.5\Delta r$, which we adopt for $\lambda_C < R_\text{thresh}$. To convert $\Phi_\text{sm}(R_\text{thresh})$ and $\vec{a}_\text{sm}(R_\text{thresh})$ to $\Phi_\text{sm}(\lambda_C)$ and $\vec{a}_\text{sm}(\lambda_C)$, we use the fact that the region within $\lambda_C$ is small and of nearly uniform density to Taylor expand $\rho$ to lowest order around the position of the test point. Putting the origin at that point,
\begin{align}
\Phi(R_\text{max}) = G \int_0^{R_\text{max}} dV \frac{\rho(\vec{r})}{r} \approx G \int_0^{R_\text{max}} dV \frac{\rho_0}{r} \propto R_\text{max}^2,
\end{align}
\begin{align}
\vec{a}(R_\text{max}) &= G \int_0^{R_\text{max}} dV \frac{\rho(\vec{r})}{r^2} \hat{r} \\ \nonumber
&\approx G \int_0^{R_\text{max}} dV \frac{\rho_0 + \vec{\partial}_r\rho|_0 \cdot \vec{r}}{r^2} \hat{r} \\ \nonumber
&= G \: \vec{\partial}_r\rho|_0 \cdot \int_0^{R_\text{max}} dV \frac{\vec{r}}{r^2} \: \hat{r} \propto R_\text{max}^2.
\end{align}
Thus both $\Phi$ and $\vec{a}$ scale with the surface area of the region which sources them.\footnote{We have verified by greatly increasing the resolution of the grid in small subvolumes that these scalings provide estimates of $\Phi_\text{ex}$ and $\vec{a}$ accurate to within 10\% for $\lambda_C$ as low as $\Delta r/20$.} For $\lambda_C < R_\text{thresh}$ we therefore set
\begin{equation}
\Phi_\text{sm}(\lambda_C) = \frac{\lambda_C^2}{R_\text{thresh}^2} \Phi_\text{sm}(R_\text{thresh}),
\end{equation}
\begin{equation}
\vec{a}_\text{sm}(\lambda_C) = \frac{\lambda_C^2}{R_\text{thresh}^2} \vec{a}_\text{sm}(R_\text{thresh}).
\end{equation}
and estimate the total (external) potential and acceleration fields as
\begin{equation}
\Phi_\text{ex} = \Phi_\text{hal} + \Phi_\text{sm}, \; \; \; \; \vec{a} = \vec{a}_\text{hal} + \vec{a}_\text{sm}.
\end{equation}
\end{enumerate}

\subsection{Monte Carlo sampling}
\label{sec:MC}

The calculations of~\cite{Desmond} used approximate maximum likelihood estimates of the input model parameters, resulting in one $\Phi_\text{ex}$ and one $\vec{a}$ field (for given $R_\text{max}$) over space. However, as anticipated in that work, it is desirable to marginalise over the uncertainties in those parameters to generate a set of plausible gravitational fields, weighted by their respective likelihoods. We implement this here by repeating our calculation of $\Phi$ and $\vec{a}_5$ $N_\text{MC}=1000$ times, in each case drawing randomly and independently from the probability distributions of the input parameters to generate a set of final fields. This generates a full probability distribution for $\Phi_\text{ex}$ and $\vec{a}_5$ at each test galaxy, and hence, by further propagation with Eq.~\ref{eq:rstar}, for the expected signal $r_*$. This will allow us to derive posteriors on $\{\Delta G, \lambda_C\}$ by means of a full likelihood framework.

Each of the three steps in Sec.~\ref{sec:maps} naturally lends itself to this approach by supplying a probability distribution for the relevant parameters, as follows:

\begin{enumerate}

\item{} AM with non-zero scatter imposes an incomplete constraint on the galaxy--halo connection, in the sense that many halos may plausibly be associated with a given galaxy. We therefore repeat the inverse-AM step 200 times to generate a set of catalogues relating 2M++ galaxies to \textsc{darksky-400} halos.\footnote{Note that we do not marginalise over the uncertainties in the AM parameters themselves, which therefore constitute a systematic uncertainty of our analysis. However, due to the tight clustering constraints on these parameters~\citep{Lehmann}, this uncertainty is almost certainly subdominant to the statistical AM uncertainty. We collect and discuss our systematic uncertainties in Sec.~\ref{sec:systematics}.}

\item{} Given values for the observational proxies $d$, $N_{10}$ and $\Phi_\text{vis,10}$, a range of correction factors $\vec{c}$ are possible, with probability distributions given by their relative frequencies over resolved \textsc{darksky-400} objects. Each random draw from their joint distribution corresponds to the association of a given real galaxy with a different simulated object at fixed $\{d,N_{10},\Phi_\text{vis,10}\}$. Thoroughly sampling this distribution therefore marginalises over the possible contributions to $\Phi$ and $\vec{a}$ from halos of mass $M_\text{vir} > 7.63 \times 10^{10} M_\odot$ hosting galaxies that are unresolved by 2M++.

\item{} Uncertainties in the inputs to the BORG algorithm -- which include galaxy number densities and bias model -- produce uncertainty in the corresponding smooth density field. Because the particle-mesh BORG run that we use produces $\lesssim10$ independent samples, we marginalise over $10$ realisations from the burnt-in part of the chain, chosen further apart from one another than the correlation length. 

\end{enumerate}

These complete the first three rows of Table~\ref{tab:params}, which lists the probability distributions of all the parameters of our model. The fourth describes an additional scatter of $10\%$ we impose between the observed and true values of halo mass, concentration and position, and the remainder will be described later in this chapter. By giving conservative uncertainties to these parameters, we intend the width of the fifth-force part of our likelihood function -- and hence our inference -- to be conservative also.

\begin{table*}
  \begin{center}
    \begin{tabular}{|l|l|l|}
      \hline
      \textbf{Parameter} & \textbf{Description / origin of uncertainty} & \textbf{Parent distribution / sampling procedure}\\
      \hline
\rule{0pt}{3ex}
      P$(M_\text{vir}, c|L; \alpha, \sigma_\text{AM})$ & Stochasticity in galaxy--halo connection & 200 mock AM catalogues at fixed $\{\alpha, \sigma_\text{AM}\}$\\
\rule{0pt}{3ex}
      P($\vec{c}|d, N_{10}, \Phi_\text{vis,10})$ & Location of unresolved halos & Random draw from frequency distribution in N-body sim.\\
\rule{0pt}{3ex}
      Smooth density field & Galaxy number densities, bias model etc & 10 independent draws from BORG posterior\\
\rule{0pt}{3ex}
      P$(M_{\text{vir},t}, c_t, \vec{r}_t|M_{\text{vir},o}, c_o, \vec{r}_o)$ &Halo properties & Uncorrelated 10\% Gaussian uncertainty\\
\rule{0pt}{3ex}
      P$(\Phi_{\text{in},t}|\Phi_{\text{in},o})$ & Internal potential & 10\% (with NSA data) or 30\% (without) Gaussian uncertainty\\
\rule{0pt}{3ex}
      P$(\text{early}|M_*)$ & Probability of being early-type & Linear fit to~\cite{Henriques}, fig. 5\\
\rule{0pt}{3ex}
      P$(M_*, R_\text{eff}, R_\text{d,gas}|M_\text{HI})$ & Galaxy properties from \textit{Alfalfa} data & Scalings of~\cite{Dutton11} with scatters as quoted there\\
\rule{0pt}{3ex}
      P$(\Sigma_{D,t}^0|\Sigma_{D,o}^0)$ & Central dynamical surface density & 0.5 (with NSA data) or 1 (without) dex Gaussian uncertainty\\
      \hline
     \end{tabular}
  \caption{Parameters forming the input to our calculation of the gas--star offset $\vec{r}_*$ for each \textit{Alfalfa} galaxy, and the distributions from which they are drawn. For each Monte Carlo realisation of the fifth-force model we sample randomly and independently from these distributions to build an empirical likelihood function $\mathcal{L}(\vec{r}_*|\lambda_C, \Delta G)$. Subscripts `t' and `o' denote `true' and `observed' values respectively. Further information is provided in Sec.~\ref{sec:method}.}
  \label{tab:params}
  \end{center}
\end{table*}

\subsection{Relating $\vec{a}$ to $\vec{a}_5$}
\label{sec:a5}

Sec.~\ref{sec:maps} describes the calculation of $\Phi$ and $\vec{a}$ from \emph{all} mass within a distance $\lambda_C$ of a given test point. $\Phi$ is ready for use in Eq.~\ref{eq:rstar}, but the acceleration of interest, $\vec{a}_5$, is due to the fifth force alone, which couples only to \emph{unscreened} mass. We must therefore determine which source masses are screened and which unscreened, in order to include only the latter. This section gives our method.

First, we calculate $\Phi_\text{ex}(\lambda_C)$ at the position of each 2M++ galaxy by the method of Sec.~\ref{sec:maps}. We estimate $\Phi_\text{in}$ by plugging into Eq.~\ref{eq:Vin} the properties of the halo associated with the galaxy in the inverse-AM step (with a random scatter given by the fifth row of Table~\ref{tab:params}). The halo is unscreened, and hence contributes to $\vec{a}_5$, if $|\Phi_\text{ex}(\lambda_C)| + |\Phi_\text{in}| < |\Phi_c|$. As $\Phi$ depends on the values of the probabilistic input parameters described in Sec.~\ref{sec:MC}, we recalculate it for each of the $N_\text{MC}$ realisations of the underlying model, and hence derive a probability $p_j(\lambda_C)$ for halo $j$ to be unscreened. We repeat this also using solely the external part of the potential (i.e. setting $\Phi_\text{in}=0$) to make $p_\text{j,ex}(\lambda_C)$, which we will use below.

With the screening properties of the 2M++ halos determined, we turn next to the remaining mass component: the smooth density field. To ascertain where this is screened, we reconstruct the complete $p_\text{ex}(\vec{r}; \lambda_C)$ field by linearly interpolating its values $p_\text{j,ex}(\lambda_C)$ at the positions of the 2M++ halos, which, by virtue of the high completeness of that survey, provide a fair sampling of the entire $z<0.05$ volume. We then simply evaluate $p_\text{ex}$ at the position $\vec{r}_k$ of each point mass $k$ representing the smooth density field, and assume that the structures in which this mass is located are sufficiently diffuse never to self-screen. $p_\text{ex}(\vec{r}_k; \lambda_C)$ thus provides the probability for portion $k$ of the smooth density field to be unscreened.

We are now in position to calculate the fifth-force acceleration $\vec{a}_5$ for each \textit{Alfalfa} galaxy. For each Monte Carlo realisation, we begin by setting each source halo and portion of the smooth density field to be screened or unscreened according to its probability $p_j(\lambda_C)$ or $p_\text{ex}(\vec{r}_k; \lambda_C)$ as calculated above.\footnote{By doing this independently for each object we assume the screening fractions to be uncorrelated. Although not strictly true, this is a good approximation: variations in most of the input parameters affect regions of size $<\lambda_C$, while typical objects are separated by much larger distances.} We calculate $\vec{a}_5(\lambda_C)$, separately for each realisation, by evaluating the sum in Eq.~\ref{eq:asum} over unscreened halos and adding the $G M/r^2$ contribution from each unscreened portion of the smooth density field within $\lambda_C$. As the screened stars do not contribute to $\vec{a}_5$ even in unscreened galaxies, however, we first subtract $M_*$ from $M_\text{vir}$ in the halo contribution. In practice this modification is minimal since stars comprise $\lesssim5$\% of total halo mass.

While calculating $\vec{a}_5$ itself requires sources at most $\lambda_\text{C,max}$ beyond the maximum distance of any test galaxy, determining whether those sources are themselves screened requires sources a further $\lambda_\text{C,max}$ out. Since the 2M++ catalogue and BORG-reconstructed density fields become unreliable beyond $\sim200$ Mpc, and, as we find below, the fixed angular uncertainty in the \textit{Alfalfa} HI-optical offset means that test points beyond $\sim100$ Mpc bring little further constraining power to $\Delta G$ and $\lambda_C$, we take $\lambda_\text{C,max} = 50$ Mpc.\footnote{By Eq.~\ref{eq:f(R)3}, $\lambda_C=50$ Mpc corresponds to $|\Phi_c| = 2.3 \times 10^{-4}$. This is larger than the potential of typical main sequence stars $\sim10^{-6}$, suggesting that in this model stars would not be sufficiently dense to screen themselves. Thus both gas and stars would feel $F_5$ and the signal $r_*$ would go away. (This model would also leave the Milky Way, with similar $\Phi$, unscreened.) Nevertheless, we include this model to explore more fully the constraining power of our signal: as in the case when screening is switched off, strictly our constraints require $F_5$ to couple differently to stars and gas.} Our minimum $\lambda_C$ value is set by the point at which the predicted $r_*$ values become so small that the corresponding $\Delta G/G_N$ constraint becomes uninteresting. We will show in Sec.~\ref{sec:constraints} that $\Delta G/G_N$ cannot be constrained to better than $\mathcal{O}(1)$ for $\lambda_C \simeq 400$ kpc, which is the value we therefore adopt for $\lambda_\text{C,min}$. This is also the approximate scale at which tidal effects within galaxies may be expected to become important.

\subsection{Calculating $\vec{r}_*$}
\label{sec:signal}

Armed with a determination of $\Phi$ and $\vec{a}_5$ for each \textit{Alfalfa} galaxy, we are now nearing our goal of using Eq.~\ref{eq:rstar} to calculate the expected signal $\vec{r}_*$ as a function of $\Delta G$ and $\lambda_C$. The remaining unknown is $M(<r_*)$, the total enclosed halo plus gas mass within the HI-optical separation $r_*$. Three classes of method present themselves for finding this: one could 1) attempt to fit halo profiles to the galaxies with empirical methods such as AM, 2) use dynamical information at small radius, or 3) employ empirical scalings between mass and light at galaxies' centres. Which of these is most precise depends on the information at one's disposal, as well as the size of the $r<r_*$ region relative to the entire galaxy or halo. We shall see that realistic $\vec{a}_5$ and interesting $\Delta G/G_N$ values within our range of $\lambda_C$ produce $r_*$ values of order 10-100 pc, at least an order of magnitude smaller than galaxy size or halo scale radius. This makes methods such as the first, which rely on global fits to galaxies' luminosity or velocity profiles, unreliable. While central kinematics would provide the most convenient and precise means of calculating $M(<r_*)$, this is not available for most \textit{Alfalfa} detections, some of which are not even clearly associated with a concentration of optical light. We are therefore forced to resort to the third method.

We estimate $M(<r_*)$ using the observed relation between central baryonic and dynamical surface mass densities reported in~\cite{Lelli_1, Lelli_2} and parametrised by~\cite{Milgrom}:
\begin{equation} \label{eq:CDR}
\Sigma_D^0 = \Sigma_M \: S(\Sigma_B^0/\Sigma_M),
\end{equation}
where
\begin{equation} \label{eq:S}
S(y) = y/2 + y^{1/2} (1+y/4)^{1/2} + 2\sinh^{-1}(y^{1/2}/2),
\end{equation}
$\Sigma_D^0$ is the central dynamical surface density, $\Sigma_B^0$ is the central baryonic surface density, and $\Sigma_M$ is a constant, $1.37 \times 10^8 M_\odot/\text{kpc}^2$. This relation captures at small $r$ the empirical trend of increasing dark matter dominance for decreasing baryonic surface density~\citep{Sanders_MDAcc, McGaugh_MDAcc, Famaey_McGaugh, Janz}: $\Sigma_D^0 \rightarrow \Sigma_B^0 + \Sigma_M$ for $\Sigma_B^0 \gg \Sigma_M$, and $\Sigma_D^0 \rightarrow (\Sigma_M \Sigma_B^0)^{1/2}$ for $\Sigma_B^0 \ll \Sigma_M$. Although possessing some scatter and hence providing a noisier determination of $\Sigma_D^0$ than would direct dynamical data, this method has the advantage of requiring neither spectroscopic information nor assumptions about the form of the overall dark matter distribution.

We take $\Sigma_D^0$ to be the projection of a core with uniform density $\rho_0$ in the very central $r<r_*$ regions of halos, an assumption  supported observationally at radii much less than the halo scale $r_s$~\citep{cusp-core, Rodrigues}. Hence
\begin{equation} \label{eq:M(r)}
M(<r_*) = \frac{4}{3} \pi \: r_*^3 \: \rho_0 = \frac{4}{3} \pi \: r_*^3 \: \frac{\Sigma_D^0 - \Sigma_*^0}{2h},
\end{equation}
where $h$ is the scale-height of the projection and we subtract $\Sigma_*^0$ from $\Sigma_D^0$ to derive the gas plus dark matter mass within the position of the stellar centroid. For realistic galaxy parameters we will find $\rho_0 \sim 10^9 \: M_\odot \: \text{kpc}^{-3}$, with a standard deviation between galaxies a factor of a few larger. Under the assumption of constant density the halo acts as a spring, with the restoring force on the stellar centroid proportional to its distance from the halo centre. As this force is required to provide a fixed acceleration equal to $\vec{a}_5$, the stiffer the spring (i.e. the greater the central density) the smaller the displacement. Plugging into Eq.~\ref{eq:rstar} and solving for $\vec{r}_*$ yields
\begin{equation} \label{eq:rstar2}
\vec{r}_* = \frac{3}{2\pi} \: \frac{\Delta G}{G_N^2} \: \frac{h}{\Sigma_D^0 - \Sigma_*^0} \: \vec{a}_5,
\end{equation}
with the following components in the right ascension (J2000; RA; $\alpha$) and declination (DEC; $\delta$) directions:
\begin{equation} \label{eq:rstarcomp}
r_{*,\alpha} = \vec{r}_* \cdot \hat{\alpha}, \; \; \; \; r_{*,\delta} = \vec{r}_* \cdot \hat{\delta}.
\end{equation}
These are the quantities we will compare between model and observations. As our observed displacements are restricted to the plane of the sky, when we speak of $\vec{r}_*$ hereafter we will typically mean $r_{*,\alpha} \hat{\alpha} + r_{*,\delta} \hat{\delta}$.

Given Eq.~\ref{eq:CDR}, our task now is to deduce $\Sigma_B^0$, the central surface density of total (star plus gas) baryonic mass. This is known to greater precision the more structural galaxy information is available, and we therefore seek to augment our knowledge of the \textit{Alfalfa} galaxies, which besides HI and OC positions have measured linewidths and total HI masses only, with more detailed optical data. For this purpose we cross-correlate the catalogue with the \textit{Nasa Sloan Atlas} (NSA),\footnote{\url{www.nsatlas.org}} a compilation of measurements and derived quantities for nearby galaxies with state of the art sky subtraction and photometric determinations~\citep{Blanton}. After cleaning the data by removing galaxies with unreliable size or redshift measurements~\citep{Bernardi}, we take from the NSA stellar mass $M_*$ (derived from {\tt kcorrect} \cite{kcorrect} with a Chabrier \cite{chabrier} initial mass function), S\'{e}rsic index \texttt{SERSIC\_N}$\:\equiv\:n$, aperture velocity dispersion $\sigma$, S\'{e}rsic half-light radius along the major axis \texttt{SERSIC\_TH50}, and apparent S\'{e}rsic minor-to-major axis ratio \texttt{SERSIC\_BA}$\:\equiv(b/a)_\text{obs}$.

We associate each \textit{Alfalfa} galaxy with the nearest NSA galaxy, provided it is within 2 Mpc. This accounts approximately for offsets in the distance estimates recorded in the two catalogues, although we caution that this cross-identification is likely to contain some errors. We show in Sec.~\ref{sec:correlations}, however, that the width of the fifth-force part of likelihood function, which measures the uncertainty in the predicted signal and is sensitive to assumptions of galaxy structure, is subdominant to that of the Gaussian component which accounts for measurement uncertainty and non-fifth-force physics. For current data, therefore, precision estimation of $\Sigma_B^0$ is not critical.

Due to the only partly overlapping footprints of the two surveys and the fact that HI and optical flux are often poorly or anti-correlated, our NSA identification succeeds for only $22$\% of \textit{Alfalfa} galaxies, and for the remainder we must use less precise means for determining $M(<r_*)$. The following subsections describe our procedure for the cases in which NSA information is or is not available.

\subsubsection{With NSA information}
\label{sec:NSA}

First, we use the observed minor-to-major axis ratio $(b/a)_\text{obs}$ to estimate the intrinsic axis ratio $(b/a)$. We do this by randomly assigning an inclination $i$ for the galaxy and setting~\cite{Haynes_84}
\begin{equation} \label{eq:axisratio}
\left(\frac{b}{a}\right)^2_\text{int} = 1 - \frac{1-(b/a)_\text{obs}^2}{\sin(i)^2}
\end{equation}
with the sole requirement that $(b/a)_\text{int} \ge 0.15$, the lowest axis ratio recorded in the NSA. We then calculate the angular diameter distance $d_A$ to each galaxy from the redshift and an assumed cosmology $\{h,\Omega_\Lambda,\Omega_m,\Omega_k\} = \{0.7, 0.7, 0.3, 0\}$ and calculate the major-axis length as $R_\text{eff}^\text{maj} \equiv d_A \times \:$\texttt{SERSIC\_TH50}. This is related to the circularised, $R_\text{eff}$, and minor-axis, $R_\text{eff}^\text{min}$, half-light radii as
\begin{equation}
R_\text{eff} = (b/a)^{1/2} \: R_\text{eff}^\text{maj} \: \: \: \: \text{and} \: \: \: \: R_\text{eff}^\text{min} = (b/a) \: R_\text{eff}^\text{maj}.
\end{equation}

Next, we analytically deproject the spherically-symmetric 3D S\'{e}rsic profile~\citep{Sersic} to derive the stellar surface density, and hence calculate the central value $\Sigma_*^0$~\citep{Prugniel, Ciotti},
\begin{equation}
\Sigma_*^0 = \frac{M_*}{2 \pi \: n \: b_n^{-2n} \: \Gamma(2n) \: R_\text{eff}^2},
\end{equation}
where $b_n \equiv 2n - 1/3 + 0.009876/n$. We then calculate the central gas density $\Sigma_g^0$. Although the total HI mass $M_\text{HI}$ is measured in \textit{Alfalfa}, the distribution of this mass is not. We take $M_\text{gas} = 1.33 \: M_\text{HI}$, to account for cosmological helium, and assume it to be distributed in an exponential disk with effective radius given by the scaling relation of~\cite{Dutton11},
\begin{equation} \label{eq:Reff}
\log_{10}(R_\text{eff,g} / \text{kpc}) = \log_{10}(0.92 \: R_\text{eff}^\text{maj} / \text{kpc}),
\end{equation}
with a Gaussian scatter of 0.25 dex. Setting $\Sigma_B^0 = \Sigma_*^0 + \Sigma_g^0$ then lets us calculate $\Sigma_D^0$ by Eq.~\ref{eq:CDR}. We estimate the scale height $h$ of the dynamical mass distribution as $R_\text{eff}^\text{min}$.

Finally, we estimate the internal potential as $\Phi_\text{in} = -\sigma^2$, with an uncertainty of 10\%.

\subsubsection{Without NSA information}
\label{sec:no-NSA}

If NSA information is not available, it is necessary to estimate not only $h$ and $R_\text{eff,g}$, but also $M_*$, $R_\text{eff}$ and the characteristic velocity that sets $\Phi_\text{in}$. We take $\Phi_\text{in}=-V_\text{max}^2$, with $V_\text{max}$ estimated from the full width half max $W_{50}$ of the radio detection
\begin{equation} \label{eq:Vmax}
V_\text{max} = \frac{W_t}{2 \sqrt{1-(b/a)^2}},
\end{equation}
where
\begin{equation} \label{eq:Wt}
W_t = \sqrt{W_{50}^2 + 4 \sigma_d^2}
\end{equation}
is a turbulence-correction linewidth~\citep{Tully}, $\sigma_d=8$ km/s is a characteristic dispersional velocity~\citep{Begum, Geha} and $(b/a)=0.2$ is the assumed intrinsic axis ratio of the system. We assign $V_\text{max}$ a 15\% Gaussian uncertainty. We then estimate the stellar properties of the galaxy using empirical scaling relations with HI properties~\citep{Dutton11}:
\begin{equation} \label{eq:Mstar}
\log_{10}(M_*/M_\odot) = 1.89 \log_{10}(M_\text{gas}/M_\odot) - 8.12
\end{equation}
with 0.3 dex scatter, and
\begin{equation} \label{eq:Reff2}
\log_{10}(R_\text{eff}^\text{maj}) = \log_{10}\left(\frac{1}{2} \: R_0 \: (M_*/M_0)^\alpha \: (1 + (M_*/M_0)^\gamma)^\frac{\beta-\alpha}{\gamma}\right)
\end{equation}
with 0.2 dex scatter. To account to first order for the size difference between ``early'' and ``late''-type galaxies, we assign each \textit{Alfalfa} galaxy a probability $P_\text{early}$ of being ``early-type'', derived by interpolating the relation derived by~\cite{Henriques} (their fig. 5) with a linear relation as a function of $\log_{10}(M_*/M_\odot)$. We use this probability to randomly set the galaxy to be early or late-type, and assign the corresponding $\{\alpha, \beta, \gamma, \log_{10}(M_0), \log_{10}(R_0)\}$ from table 1 of~\cite{Dutton11}. We then use Eq.~\ref{eq:Reff} to estimate $R_\text{eff,g}$, and proceed as in Sec.~\ref{sec:NSA}.

\subsubsection{Sampling and scatter in $\Sigma^0_D$}
\label{sec:more-sampling}

As in Sec.~\ref{sec:MC}, we sample each probability distribution in Secs.~\ref{sec:NSA} and~\ref{sec:no-NSA} randomly and independently for each Monte Carlo realisation of our model, effectively marginalising over them in our determination of the likelihood distribution of $\vec{r}_*$. This fills in rows $6$ and $7$ of Table~\ref{tab:params}.

Besides having relatively large scatter, however, these scalings may be expected to provide good fits only to subsets of the total galaxy population, and there is no guarantee that they will accurately describe the \textit{Alfalfa} sample. In order to ensure that our constraints on $\Delta G$ are nevertheless conservative, we therefore scatter $\Sigma^0_D$ by an additional $0.5$ dex in the case where NSA is available, and $1$ dex when it is not (final row of Table~\ref{tab:params}). This propagates directly into $r_{*,\alpha}$ and $r_{*,\delta}$ by Eq.~\ref{eq:rstar2}, and guarantees that the galaxies with NSA information are prioritised in the inference. We find that increasing the scatter of $\Sigma^0_D$ in the model increases predicted $r_*$ values on the whole, and hence reduces inferred values of $\Delta G/G_N$. Given that this scatter is itself significantly uncertain, our results for $\Delta G/G_N$ in Sec.~\ref{sec:constraints} ought not be considered more than order of magnitude estimates. In Sec.~\ref{sec:systematics} we will show explicitly the result of implementing different scatters for $\Sigma^0_D$. On the other hand, global variations in $\Sigma^0_D$ scatter are scale-independent and hence introduce no bias in the $\lambda_C$-dependence of any quantity.

Ideally, unknowns such as this would be treated as free parameters in the inference and marginalised over in the determination of $\Delta G/G_N$ and $\lambda_C$. This may become possible with data of greater quality and quantity in the future.

\subsection{Modelling the noise}
\label{sec:uncertainties}

The constraining power of the gas--star offsets for fifth-force parameters depends crucially on the angular measurement uncertainty $\theta$. As the OC centroid is typically known to 3 arcsec or better, this uncertainty is dominated by the 21 cm radio detection of HI. This depends on the telescope pointing, data reduction pipeline and signal to noise ratio $s$ of the measurement, and is not therefore well known a priori. We will determine it empirically from the data itself. Our fiducial method is to set $\theta_\alpha$ and $\theta_\delta$ equal to the dispersion in the RA and DEC components of the measured HI-OC offset itself in bins of $\log_{10}(s)$, as recommended by \cite{Haynes}. In particular, we create 50 uniform bins between the minimum and maximum values $0.66$ and $3$, and in each bin calculate separately the standard deviation of $r_{*,\alpha}$ and $r_{*,\delta}$. These have a mean of 18 arcsec over the entire sample and a standard deviation among galaxies of 4 arcsec.

This method guarantees that the measured offsets will be consistent with $0$ to $\sim1\sigma$, and hence requires that any potential fifth-force contribution to the signal be relatively small. Although we will show this to be the case in Sec.~\ref{sec:correlations} for all reasonable values of $\Delta G/G_N$, a more self-consistent approach is to parametrise the HI-OC uncertainties and fit for them simultaneously with $\Delta G$ and $\lambda_C$. We present three variants of this approach, each assuming that $\theta$ is correlated only with $s$. In the first, we assume a universal value $\bar{\theta}$ for both $\theta_\alpha$ and $\theta_\delta$, and fit for $\{\Delta G, \lambda_C, \bar{\theta}\}$. In the second and third we model $\theta$ as a polynomial in $\log_{10}(s)$: in one case we fit
\begin{equation} \label{eq:unc_shorter}
\theta = A + B \log_{10}(s),
\end{equation}
taking $A$ and $B$ as free parameters, and in the other
\begin{equation} \label{eq:unc_short2}
\theta = C + D \log_{10}(s) + E \log_{10}(s)^2,
\end{equation}
for $\log_{10}(s) < 1.6$, and $\theta = F$ otherwise, for $C$, $D$, $E$ and $F$ free. Each of $A-F$ are given flat priors, with $A,C,F > 0$ to force $\theta>0$. We will show in Sec.~\ref{sec:uncertainty_model} that in each case $\theta$ converges to values consistent with those of the first method and produces an almost identical $\Delta G/G_N$ posterior. This is because by far the greater part of the signal derives from noise for the $\Delta G/G_N$ values of interest in our analysis. Marginalising over $\bar{\theta}$, $A-B$ or $C-F$ broadens the $\Delta G/G_N$ posteriors only slightly. As discussed further in Sec.~\ref{sec:systematics}, operationally $\theta$ also includes the contribution to the signal from other physics, under the assumption that it supplies a Gaussian component to the likelihood with width a function only of $s$. The degeneracy between measurement uncertainty and additional physics in the determination of $\theta$ will become important for future HI surveys with greater spatial resolution (see Sec.~\ref{sec:forecast}).

We will also show results for a choice we call ``conservative,'' in which the $\theta$ values of the fiducial method are doubled. This yields robust upper limits on $\Delta G/G_N$, and is the choice we focused on in~\cite{Desmond_PRL}. Note however that this model produces significantly lower overall likelihood values than the fiducial one due to overprediction of the dispersions of the HI-OC displacements at each $s$, and in a model comparison sense is therefore clearly disfavoured.

\subsection{Likelihood model}
\label{sec:likelihood}

Over the $N_\text{MC}$ realisations of our probabilistic model, and for given $\Delta G$ and $\lambda_C$, we have now derived a distribution of predicted $r_{*,\alpha}$ and $r_{*,\delta}$ values for each \textit{Alfalfa} galaxy. Heuristically these constitute the likelihood distribution of the signal, and by scoring it against the observations and applying Bayes' theorem one can derive constraints on $\Delta G$ and $\lambda_C$ by MCMC. In practice we proceed as follows.

We begin by constructing a template signal, assuming that each test galaxy is unscreened and taking $\Delta G/G_N=1$ and one of 20 values for $\log_{10}(\lambda_C/\text{Mpc})$ uniformly spaced between $\log_{10}(0.4)$ and $\log_{10}(50)$. For each test galaxy, we find the minimum ($r_0$) and maximum ($r_1$) of $r_{*,\alpha}$ and $r_{*,\delta}$ separately over the $N_\text{MC}$ realisations of the model. We then partition the results of these realisations into $N_\text{bins}=10$ uniform bins between these limits and define $P_j \equiv N_j/N_\text{MC}$, where $N_j$ is the number of realisations falling in bin $j$. We take this histogram to represent the unscreened component of the model's likelihood function in $\{r_{*,\alpha}, r_{*,\delta}\}$ space. We calculate also the fraction $f$ of realisations in which the galaxy in question is unscreened. Given that the model predicts $r_*=0$ for the screened remainder, and treating the orthogonal RA and DEC components as independent, the overall likelihood $\mathcal{L}_i(r_{*,\alpha}, r_{*,\delta})$ for test galaxy $i$ to have $\vec{r}_*$ components $r_{*,\alpha}$ and $r_{*,\delta}$ is given by
\begin{equation} \label{eq:L1}
\mathcal{L}_i(r_{*,\alpha}, r_{*,\delta}|\Delta G, \lambda_C) = \mathcal{L}_i(r_{*,\alpha}|\Delta G, \lambda_C) \times \mathcal{L}_i(r_{*,\delta}|\Delta G, \lambda_C),
\end{equation}
where
\begin{align} \label{eq:L2}
\mathcal{L}_i(r_{*,\alpha}|&\Delta G, \lambda_C) = (1-f) \: \delta(r_{*,\alpha}) \\ \nonumber
&+ f \: \Sigma_{j=0}^{N_\text{bins}} \: P_j \: \delta(r_{*,\alpha} - \Delta G \: (r_{0,\alpha} + j \Delta r_\alpha)).
\end{align}
Here $\Delta r_\alpha \equiv (r_{1,\alpha} - r_{0,\alpha})/(N_\text{bins}-1)$ is the width of each bin, and we have approximated the unscreened component of the likelihood as a discrete sum of delta-functions $\delta$ at the midpoints of the bins. $f$, $P_j$, $r_0$ and $r_1$ are implicit functions of $\lambda_C$: by Eq.~\ref{eq:f(R)3} and the procedure of Sec.~\ref{sec:method}, a larger $\lambda_C$ corresponds to a larger $|\Phi_c|$ and hence a larger unscreened fraction $f$ and signal limits $r_0$ and $r_1$. Since $r_*$ is proportional to $\Delta G$ (Eq~\ref{eq:rstar}), this parameter simply multiplies the bin positions. The likelihood for the DEC component is defined analogously.

We convolve this with a Gaussian probability distribution function for these \emph{true} values given \emph{observed} values $r_{*,\alpha,\text{obs}}$ and $r_{*,\delta,\text{obs}}$ and their measurement uncertainties $\sigma_i$. Dropping the dependence on $\Delta G$ and $\lambda_C$ for brevity:
\begin{equation} \label{eq:L3}
\mathcal{L}_i(r_{*,\alpha,\text{obs}}) = \int dr_{*,\alpha} \; \mathcal{L}_i(r_{*,\alpha}) \: \times \: \mathcal{L}_i(r_{*,\alpha, \text{obs}}|r_{*,\alpha}),
\end{equation}
where
\begin{equation} \label{eq:L4}
\mathcal{L}_i(r_{*,\alpha, \text{obs}}|r_{*,\alpha}) = \frac{\exp\{{-(r_{*,\alpha, \text{obs}} - r_{*,\alpha})^2}/(2 \sigma_{i,\alpha}^2)\}}{\sqrt{2 \pi} \: \sigma_{i,\alpha}}.
\end{equation}
The observational uncertainty is given by
\begin{equation} \label{eq:ra_unc}
\sigma_{i,\alpha} = d_{A,i} \: \cos(\delta_i) \: \theta_i,
\end{equation}
where $d_{A,i}$ is the angular diameter distance to galaxy $i$, $\delta_i$ is the galaxy's declination, and $\theta_i$ is the angular uncertainty assigned by either the fiducial or conservative method of Sec.~\ref{sec:data}.\footnote{As explained in Secs.~\ref{sec:uncertainties} and~\ref{sec:forecast}, $\theta$ also models a Gaussian component of the likelihood model due to non-fifth-force physics. We refer to it as measurement uncertainty here simply for brevity.} The corresponding DEC uncertainty is given by
\begin{equation} \label{eq:dec_unc}
\sigma_{i,\delta} = d_{A,i} \: \theta_i.
\end{equation}
Performing the convolution in Eq.~\ref{eq:L3} then yields
\begin{align} \label{eq:L7}
&\mathcal{L}_i(r_{*,\alpha,\text{obs}}|\Delta G, \lambda_C) = (1-f) \: \frac{\exp\{{-r_{*,\alpha, \text{obs}}^2}/(2\sigma_{i,\alpha}^2)\}}{\sqrt{2 \pi} \: \sigma_{i,\alpha}} \\ \nonumber
& + f \: \Sigma_{j=0}^{N_\text{bins}} \: P_j \: \frac{\exp\{{-(r_{*,\alpha,\text{obs}} - \Delta G \: r_j)^2/(2\sigma_{i,\alpha}^2)}\}}{\sqrt{2 \pi} \: \sigma_{i,\alpha}},
\end{align}
where $r_j \equiv (r_{0,\alpha} + j \Delta r_\alpha)$ is the position of the $j$th bin centre. Treating the galaxies as independent, the likelihood of the entire dataset $d$ is then
\begin{equation}
\mathcal{L}(d|\Delta G, \lambda_C) = \prod_i \mathcal{L}_i(r_{*,\alpha,\text{obs}}|\Delta G, \lambda_C) \: \mathcal{L}_i(r_{*,\delta,\text{obs}}|\Delta G, \lambda_C).
\end{equation}
Finally, we apply Bayes' theorem to deduce the probability of $\Delta G$ and $\lambda_C$ themselves:
\begin{equation}
P(\Delta G, \lambda_C|d) = \frac{\mathcal{L}(d|\Delta G, \lambda_C) \: P(\Delta G, \lambda_C)}{P(d)}.
\end{equation}
where $P(\Delta G, \lambda_C)$ is the prior distribution of the model parameters and $P(d)$ is the constant probability of the data for any $\{\Delta G, \lambda_C\}$. We take a uniform prior in $\log_{10}(\lambda_C/\text{Mpc})$ in the range $\log_{10}(0.4)-\log_{10}(50)$, and the improper prior $\Delta G \ge 0$ flat in $\Delta G$.\footnote{We find a Jeffrey's prior to yield almost identical results.} We will find our likelihood function to provide clear upper bounds on $\Delta G$, obviating the need to impose such a bound a priori.

Deriving the likelihood function directly from the results of the Monte Carlo realisations in this way removes the need to assume a particular functional form for it. This reduces bias in the inference, and we have checked explicitly that our method accurately reconstructs $\Delta G/G_N$ when mock data generated with a particular value is fed in (Sec.~\ref{sec:valid}). Since several of the most important input probability distributions in Table~\ref{tab:params} are non-Gaussian, we find that assuming the overall likelihood to be Gaussian (i.e. using the mean and standard deviation of the $N_\text{MC}$ predicted $r_{*,\alpha}$ and $r_{*,\delta}$ values rather than the full histogrammed distribution) biases inference of a known input $\Delta G/G_N$ by a factor of $\sim2$.

\subsection{Convergence and consistency tests}
\label{sec:convergence}

We have taken 10 realisations of the BORG smooth density field and 200 of the AM galaxy--halo connection. We verify that these numbers are sufficient to fully propagate these probability distributions into the predicted signal by directly investigating the impact of changing them. While with fewer realisations the spread in $r_{*,\alpha}$ and $r_{*,\delta}$ in the signal templates rises with the number of realisations, indicating additional uncertainty in the marginalisation, this effect levels off near the values chosen. Using a larger number of realisations would not therefore appreciably impact the results. We have also checked that using a different set of 10 smooth density fields and 200 AM galaxy--halo connections does not alter any of our conclusions.

Similarly, we have used a finite number of Monte Carlo draws from the model to estimate the likelihood function. With $N_\text{MC} \lesssim 1000$, repetitions of the entire procedure yield constraints on $\Delta G/G_N$ and $\lambda_C$ (Sec.~\ref{sec:results}) that differ at the $\gtrsim 20$\% level. That this is not true with $\gtrsim 1000$ realisations indicates that in this case the probability distributions of the input parameters of Table~\ref{tab:params} have been thoroughly sampled and their uncertainties fully marginalised over in the final constraints. We check also that varying our estimates of $\Phi_\text{in}$, $M_\text{vir}$, $c$, $\Sigma_D^0$, $M_*$, $R_\text{eff}$ and $R_\text{d,gas}$ within a factor of $2$ does not significantly affect the results; this change is subdominant not only to the measurement uncertainties, but also to the other theoretical uncertainties deriving from the smooth density field, galaxy--halo connection and location of low-mass halos. (The \emph{scatter} in $\Sigma_D^0$ is however important; see Sec.~\ref{sec:systematics}.)

We check that our inference is not driven by potential outliers by excising up to 10\% of points with the most extreme values of $\langle r_{*,\alpha} \rangle$, $\langle r_{*,\delta} \rangle$, $r_{*,\alpha,\text{obs}}$ or $r_{*,\delta,\text{obs}}$, where angle brackets denote a modal average over the $N_\text{MC}$ model realisations. This has little impact on the results. In Sec.~\ref{sec:valid} we will show explicitly that jackknife or bootstrap resamples of the \textit{Alfalfa} dataset behave similarly.

Finally, we check the convergence of our MCMC chains (run with the affine-invariant Python sampler \textsc{emcee}) by visual inspection and by means of the Gelman-Rubin statistic. The low dimensionality of the inference combined with the relatively simple form of the likelihood function and unimodality of the posteriors makes exploring the parameter space relatively easy. With 10 walkers, we find 300 burn-in steps to suffice for 1D inference ($\Delta G/G_N$ at fixed $\lambda_C$), and 600 burn-in steps to suffice for 2D inference (both $\Delta G/G_N$ and $\lambda_C$).

We caution that errors in the probability distributions of the model parameters (including the implicit choice of delta-function priors for parameters which should in reality have some uncertainty) may lead to systematic error. We discuss this further in Sec.~\ref{sec:systematics}.

\section{Results}
\label{sec:results}

\subsection{Comparison of observed and predicted signal}
\label{sec:correlations}

\begin{figure*}
  \subfigure[]
  {
    \includegraphics[width=0.48\textwidth]{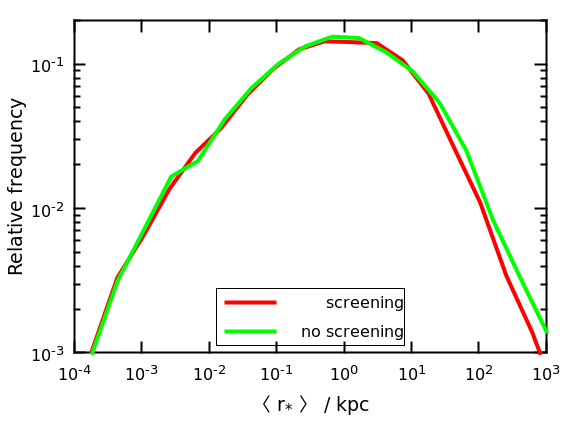}
    \label{fig:1a}
  }
  \subfigure[]
  {
    \includegraphics[width=0.48\textwidth]{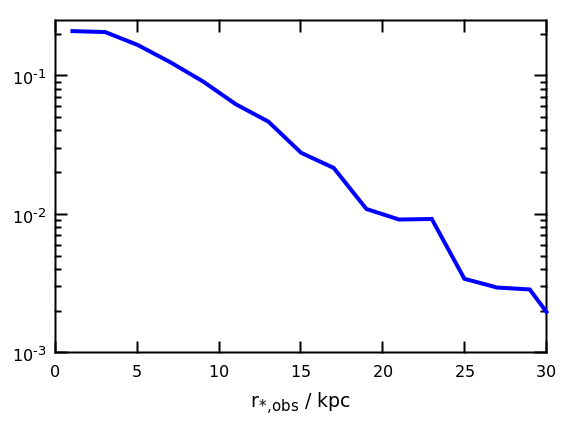}
    \label{fig:1b}
  }
  \subfigure[]
  {
    \includegraphics[width=0.48\textwidth]{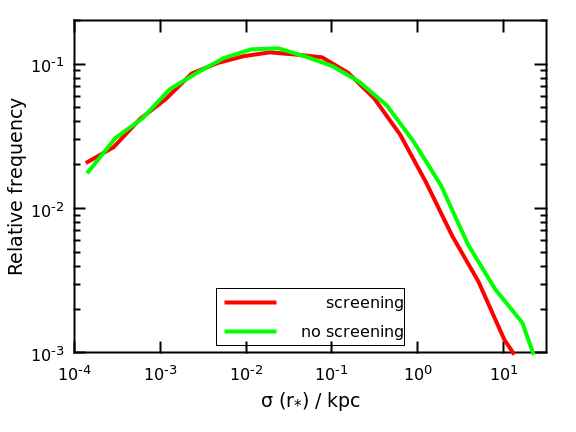}
    \label{fig:1c}
  }
  \subfigure[]
  {
    \includegraphics[width=0.48\textwidth]{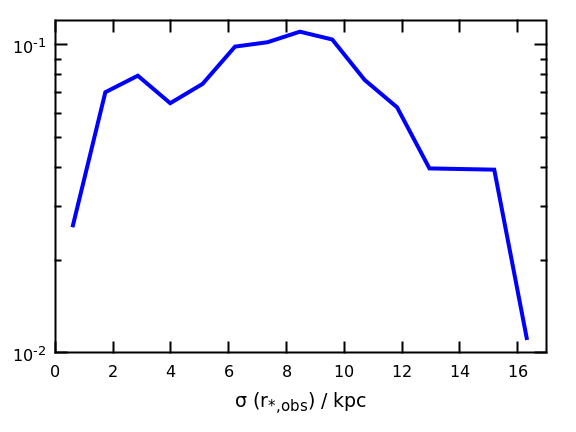}
    \label{fig:1d}
  }
  \caption{Histograms of the expectation (upper) and uncertainty (lower) of the fifth-force predicted (left) and observed (right) gas--star offset $r_*$ over all \textit{Alfalfa} galaxies. For the fifth-force prediction we take the mode of the 1000 Monte Carlo realisations for each galaxy of the screened model with $\lambda_C = 5$ Mpc and $\Delta G/G_N = 1$, given 10 bins for both the RA and DEC components ($r_{*,\alpha}$ and $r_{*,\delta}$) between their minimum and maximum values. The uncertainty is the minimal width enclosing $68\%$ of the realisations. For the observed signal the uncertainty is assigned using the fiducial method of Sec.~\ref{sec:uncertainties}. $r_{*,\text{obs}}$ is typically several times larger than $\langle r_* \rangle$, and the corresponding uncertainty two orders of magnitude larger, even for the relatively extreme case in which the fifth force is as strong as gravity ($\Delta G = G_N$). The majority of the signal must therefore be due to non-fifth-force effects.}
  \label{fig:hists}
\end{figure*}

To begin our investigation of the presence of a fifth-force signal in the \textit{Alfalfa} data, we show in Fig.~\ref{fig:1a} (red) the distribution of the predicted values of $\langle r_* \rangle = \langle (r_{*,\alpha}^2 + r_{*,\delta}^2)^{1/2} \rangle$ for fiducial model parameters $\Delta G/G_N=1$ and $\lambda_C=5$ Mpc. For each galaxy this is the mode of the binned distribution of the $N_\text{MC}$ model realisations, i.e. $\langle r_* \rangle \equiv r_0 + j_0 \Delta r$ where $j_0$ maximises $P_j$. For comparison, we show in green the case in which screening is switched off. This is tantamount to setting $|\Phi_c| \rightarrow \infty$, implying $f=1$ for all \textit{Alfalfa} galaxies and that all matter within $\lambda_C$ of each test point is fully unscreened. Thus any pair of masses within that separation interact with a fifth force, leading to an enhanced signal. We show the analogous distribution for the real data in Fig.~\ref{fig:1b}. In Fig.~\ref{fig:1c} we show the uncertainty on the predicted $r_*$, defined as the minimal width enclosing 68\% of the model realisations. As in Fig.~\ref{fig:1a} we show the results both with and without screening. Fig. ~\ref{fig:1d} shows the distribution of uncertainties $\theta_i$ in the HI centroids in the real data, assigned by the fiducial method.

\begin{figure*}
  \subfigure[Predicted $r_*-\Phi$]
  {
    \includegraphics[width=0.48\textwidth]{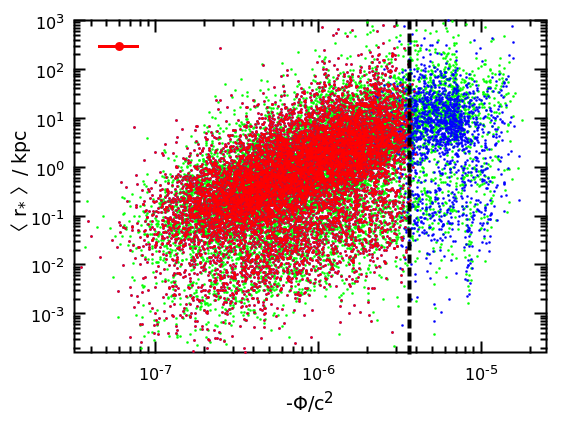}
    \label{fig:2a}
  }
  \subfigure[Predicted $r_*-a_5$]
  {
    \includegraphics[width=0.48\textwidth]{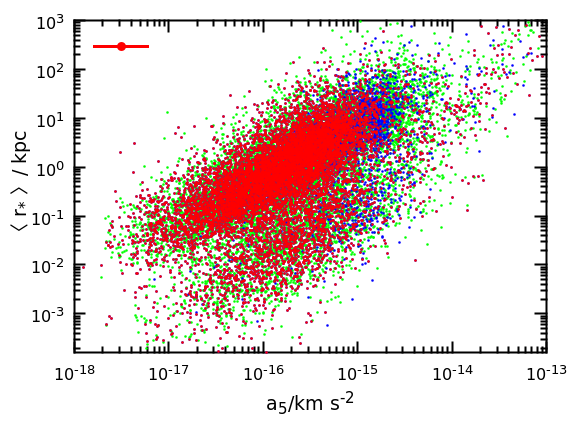}
    \label{fig:2b}
  }
  \subfigure[Observed $r_*-\Phi$]
  {
    \includegraphics[width=0.48\textwidth]{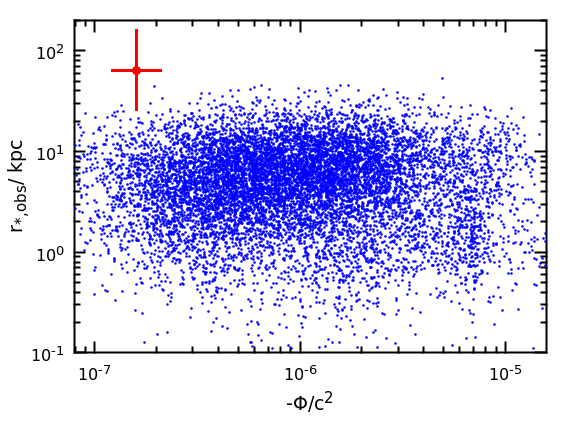}
    \label{fig:2c}
  }
  \subfigure[Observed $r_*-a_5$]
  {
    \includegraphics[width=0.48\textwidth]{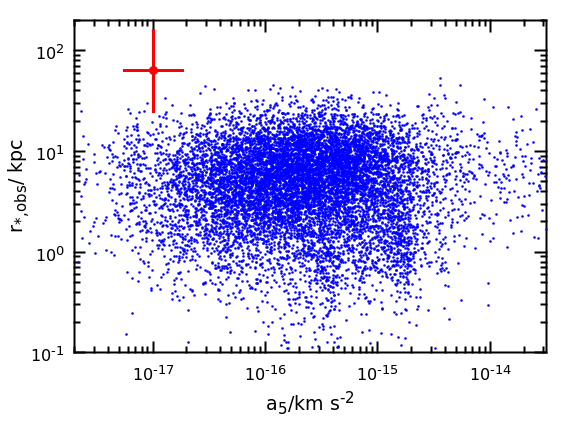}
    \label{fig:2d}
  }
  \caption{Correlations of predicted (upper) and observed (lower) signal with Newtonian potential $\Phi$ (left) and magnitude of fifth-force acceleration $a_5$ (right). The prediction is calculated as in Fig.~\ref{fig:hists}, again for the case $\lambda_C = 5$ Mpc, $\Delta G/G_N=1$, and $\langle \Phi \rangle$ and $\langle a_5 \rangle$ are the modal averages over the 1000 Monte Carlo model realisations. In the top row, the red points are for the full chameleon- or symmetron-screened model, the blue points as the red except without the $r_*$ values of screened galaxies set to 0, and the green points show the case where screening is entirely switched off. The black dashed line shows the threshold $|\Phi_c|$, above which galaxies are screened. The median sizes of the errorbars in the four directions -- minimal 68\% bounds for the prediction, $1\sigma$ for the observations -- are shown by the red crosses in the top left corners (the uncertainties in $\langle r_* \rangle$ are the same size as the red point). While the predictions show a clear cutoff in $r_*$ for $|\Phi|>|\Phi_c|$ and a positive correlation with $a_5$ (Eq.~\ref{eq:rstar}), this is not apparent in the observations.}
  \label{fig:Va_corr}  
\end{figure*}

\begin{figure*}
  \subfigure[RA projection]
  {
    \includegraphics[width=0.48\textwidth]{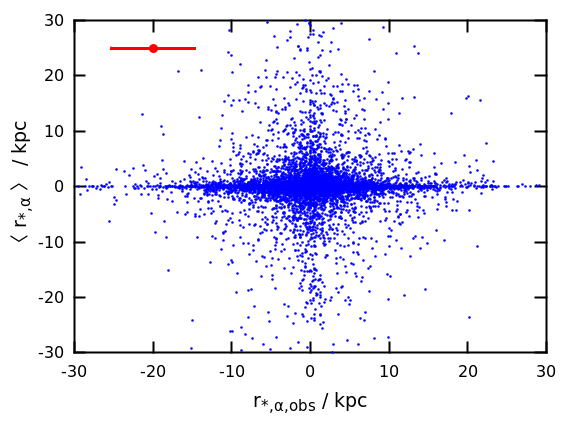}
    \label{fig:3a}
  }
  \subfigure[DEC projection]
  {
    \includegraphics[width=0.48\textwidth]{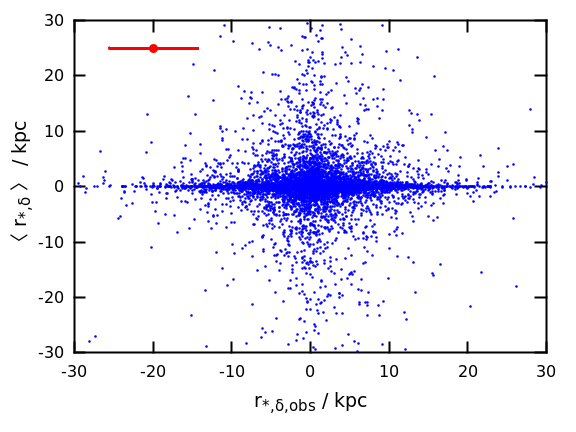}
    \label{fig:3b}
  }
  \caption{Correlation of the observed and predicted components of $r_*$ in the RA ($r_{*,\alpha}$, left) and DEC ($r_{*,\delta}$, right) directions, with the predicted signal again calculated as in Fig.~\ref{fig:hists} and errorbars as in Fig.~\ref{fig:Va_corr}.}
  \label{fig:radec_corr}
\end{figure*}

Fig.~\ref{fig:2a} shows the correlation of $\langle r_* \rangle$ with $\Phi$. The green points are for the case in which screening is switched off, the red points for when it is included -- with the signal of screened ($f<0.5$) galaxies set to 0 -- and the blue points as the red except without the $f<0.5$ points being lowered. The black dashed line shows the screening threshold $\Phi_c$ for this choice of $\lambda_C$, and the median errorbars in both directions, enclosing 68\% of the model realisations, are shown in the top left. Fig.~\ref{fig:2b} shows the analogous correlation with $a_5$ with identical colour scheme, and Figs.~\ref{fig:2c} and~\ref{fig:2d} show the corresponding correlations for the real data. In Figs.~\ref{fig:3a} and~\ref{fig:3b} we plot the RA and DEC components of the observed and predicted signals directly against each other.

We draw four inferences at this stage:

\begin{enumerate}

\item{} For this choice of model parameters, the predicted signals are typically $\mathcal{O}(1)$ kpc while the observed ones are $\mathcal{O}(10)$ kpc with a narrower distribution. This reflects the strength of the restoring force on the stellar centroid due to its offset from the halo centre (derived from the central density in Eq.~\ref{eq:M(r)}), on the one hand, and the relatively large observational uncertainties in the HI centroid position on the other. This indicates that were the observed $r_*$ values due entirely to a fifth force, the required $\Delta G/G_N$ would be $\mathcal{O}(10)$. If competitive constraints are to be placed on $\Delta G$ with this signal, therefore, it must primarily be by means of the correlation of the relative directions and magnitudes of $\vec{r}_*$ and $\vec{a}_5$ across the galaxy sample. Assuming $\Delta G/G_N \lesssim 1$, modified gravity can account for at most $\sim10\%$ of the HI-OC offset.

\item{} For $\Delta G/G_N=1$, the uncertainty on the predicted signal (width of the likelihood function) is typically $\sim 10^{-2}-10^{-1}$ kpc, while that of the observations is again $\mathcal{O}(10)$ kpc. As these two terms contribute equally to the likelihood in Eq.~\ref{eq:L7}, this implies that the strength of the constraints we set in Sec.~\ref{sec:constraints} is determined predominantly by the precision $\theta$ with which the HI centroid of the \textit{Alfalfa} galaxies is located. As the width of the likelihood function scales with $\Delta G$ (Eq.~\ref{eq:L7}), this is even more pronounced at the small $\Delta G/G_N$ values ($\mathcal{O}(10^{-2})$) of interest there.

\item{} Both $\langle r_* \rangle$ and $\sigma(r_*)$ are slightly larger for the unscreened than screened model. This is because $\vec{a}_5$ is larger in the unscreened case due to contributions from additional masses in the vicinity of a given test point. Note however that this does not take into account the fact that in the model with screening the test galaxies themselves may be screened and hence have a predicted $r_*$ of 0 (the results here are only for the unscreened fraction). In terms of our inference this will prove to be the most significant difference between the two models.

\item{} By construction $\langle r_* \rangle$ shows clear and characteristic dependence on $\Phi$ and $a_5$. The predicted signal is proportional to $a_5$ by Eq.~\ref{eq:rstar}, generating a positive correlation whether or not screening is included. Without screening, the positive correlation between $|\Phi|$ and $a_5$ (which are sourced by the same mass distribution) simply induces a positive correlation also between $r_*$ and $|\Phi|$. The introduction of screening has two effects. First, some source masses surrounding test galaxies in high-density regions become screened and hence no longer contribute to $\vec{a}_5$, causing the predicted signal to drop: this is the difference between the blue and green points at $|\Phi|>|\Phi_c|$ in Fig.~\ref{fig:2a}. That this offset is small indicates that in this regime $\Phi$ is set mostly by the test galaxies' own masses rather than their environment, so that most surrounding mass is unscreened and hence contributes to $a_5$ and $r_*$ even when screening is included. Second, test galaxies with $|\Phi|>|\Phi_c|$ become themselves screened, causing their predicted signal to fall to 0 (red). These galaxies may span the entire range of $a_5$ (Fig.~\ref{fig:2b}), although are more likely to be found at higher values; the largest predicted signals in the case with screening therefore tend to be produced in galaxies very close to being screened. On the contrary, $r_{*,\text{obs}}$ shows no apparent correlation with $\Phi$ or $a_5$. This again implies that the greater part of the observed signal must be of non-fifth-force origin.

\end{enumerate}

These estimations of fifth-force effect are relatively coarse. A precise quantitative determination of the preference of the data for $\Delta G > 0$ -- and of the constraints on $\Delta G$ and $\lambda_C$ that it implies -- requires the full likelihood formalism of Sec.~\ref{sec:method}. It is to this that we now turn.

\subsection{Constraints on $\Delta G$ and $\lambda_C$}
\label{sec:constraints}

\subsubsection{Effect of noise model}
\label{sec:uncertainty_model}

\begin{figure*}
  \subfigure[]
  {
    \includegraphics[width=0.315\textwidth]{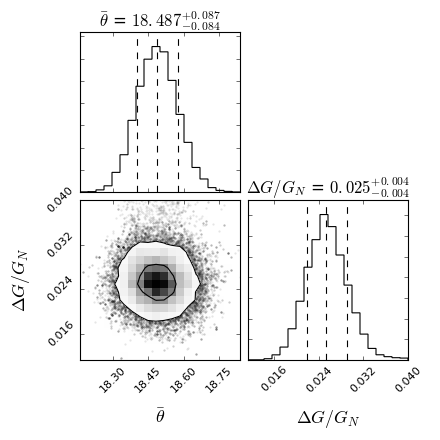}
    \label{fig:corner1}
  }
  \subfigure[]
  {
    \includegraphics[width=0.315\textwidth]{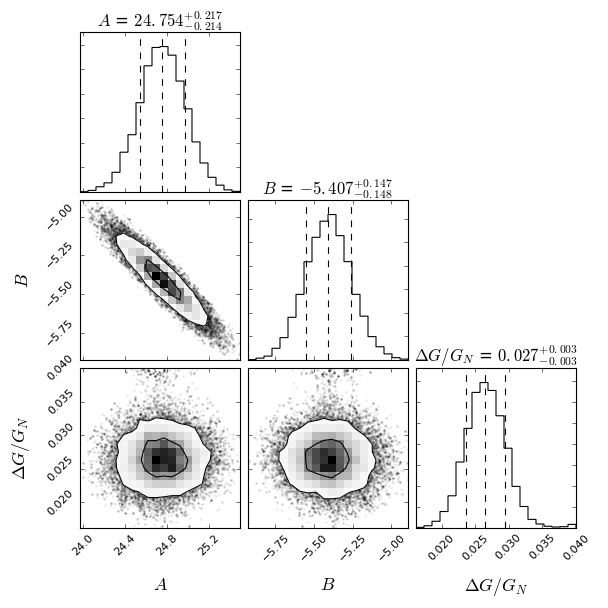}
    \label{fig:corner2}
  }
  \subfigure[]
  {
    \includegraphics[width=0.315\textwidth]{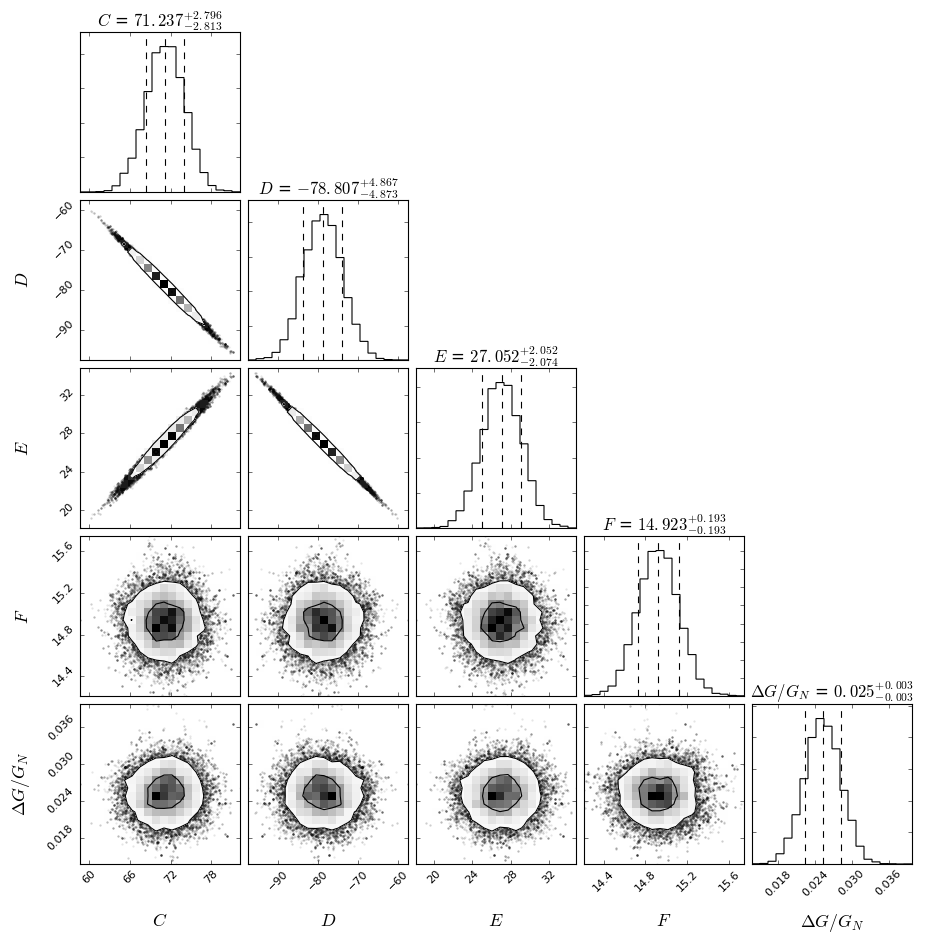}
    \label{fig:corner3}
  }
  \caption{For the screened model with $\lambda_C = 1.8$ Mpc, posteriors of $\Delta G/G_N$ and additional parameters specifying the angular width $\theta_i$ of the Gaussian, non-fifth-force component of the likelihood function for galaxy $i$. In Fig.~\ref{fig:corner1}, $\theta_i = \bar{\theta}$ arcsec for all galaxies. In Figs.~\ref{fig:corner2} and~\ref{fig:corner3} $\theta_i$ is determined as a function of signal to noise ratio $s$ of the detection according to Eqs. \ref{eq:unc_shorter} and~\ref{eq:unc_short2} respectively. In each case the chains converge to the same solution for $\Delta G/G_N$ as when $\theta$ is set a priori to the standard deviation of the measured $r_*$ in bins of $\log_{10}(s)$ (Fig.~\ref{fig:dG3}), indicating that our inference is robust to variations in the noise model.}
  \label{fig:corner}
\end{figure*}

\begin{figure}
  \centering
  \includegraphics[width=0.5\textwidth]{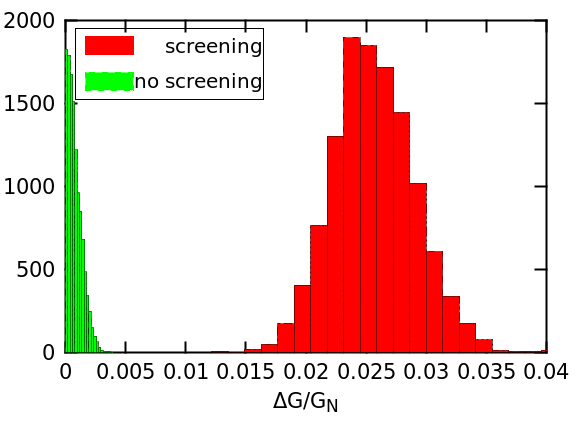}
  \caption{Posterior of $\Delta G/G_N$ for $\lambda_C = 1.8$ Mpc under the fiducial noise model, with and without screening.}
  \label{fig:dG3}
\end{figure}

From Sec.~\ref{sec:correlations} it is clear that for any reasonable value of $\Delta G/G_N$ modified gravity accounts for at most a small fraction of the total gas--star offset, justifying the fiducial method of Sec.~\ref{sec:uncertainties} for setting the measurement uncertainties $\theta_i$. This ought to manifest in an agreement between the results of that model and those of the others in that section where $\theta$ is parametrised and fitted for. Here we show this explicitly for the screened model with $\lambda_C = 1.8$ Mpc.\footnote{We focus on this value of $\lambda_C$ because it maximises the overall likelihood, as will be shown in Sec.~\ref{sec:screening}.} Fig.~\ref{fig:corner} shows the posteriors of $\{\Delta G/G_N, \bar{\theta}\}$ in the case where $\theta = \bar{\theta}$ arcsec is a free parameter, and $\{\Delta G/G_N, A, B\}$ and $\{\Delta G/G_N, C, D, E, F\}$ for the models of Eqs.\ref{eq:unc_shorter} and~\ref{eq:unc_short2} respectively. These are to be compared with Fig.~\ref{fig:dG3}, which shows the $\Delta G/G_N$ posterior for the fiducial noise model in which $\theta$ is fixed a priori to equal the standard deviation of $r_{*,\text{obs}}$ in bins of $\log_{10}(s)$. The same solution for $\Delta G/G_N$ is picked out in each case, and in Fig.~\ref{fig:corner1} it is apparent that the maximum-likelihood $\bar{\theta}$ is close to the average scatter in $r_{*,\text{obs}}$ across the entire dataset (18 arcsec). The model of Eq.~\ref{eq:unc_short2} picks out almost the same $\theta(s)$ as \cite{Haynes} eq. 1. We have checked that these results hold at other $\lambda_C$ values also. For simplicity we therefore restrict ourselves from now on to the fiducial model in which $\theta$ is fixed from the outset.

\subsubsection{1D inference without screening}
\label{sec:noscreening}

\begin{figure}
  \centering
  \includegraphics[width=0.5\textwidth]{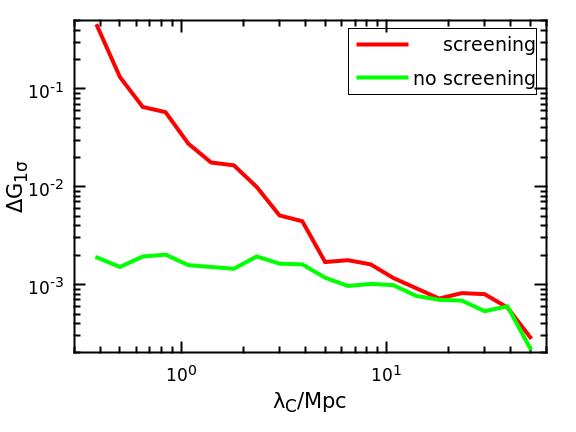}
  \caption{$1\sigma$ upper limit on $\Delta G/G_N$ $(\Delta G_{1\sigma}$), as a function of $\lambda_C$, using the fiducial uncertainties for the model without screening (green) and the conservative uncertainties for the model with screening (red). Switching to the conservative uncertainties makes only a small ($\sim0.1$ dex) difference in the unscreened case. The results with screening are weaker at small $\lambda_C$ first because a sizeable fraction of the test galaxies are screened, and hence have predicted $r_*=0$, and second because fewer neighbouring masses contribute to $\vec{a}_5$.}
  \label{fig:constraint}
\end{figure}

\begin{figure*}
  \subfigure[$\lambda_C = 500$ kpc]
  {
    \includegraphics[width=0.3\textwidth]{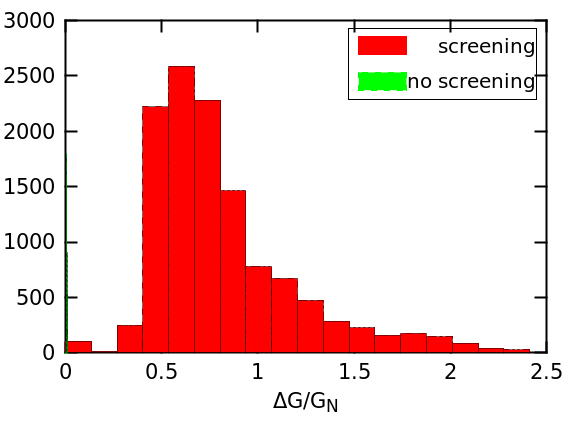}
    \label{fig:6a}
  }
  \subfigure[$\lambda_C = 5$ Mpc]
  {
    \includegraphics[width=0.3\textwidth]{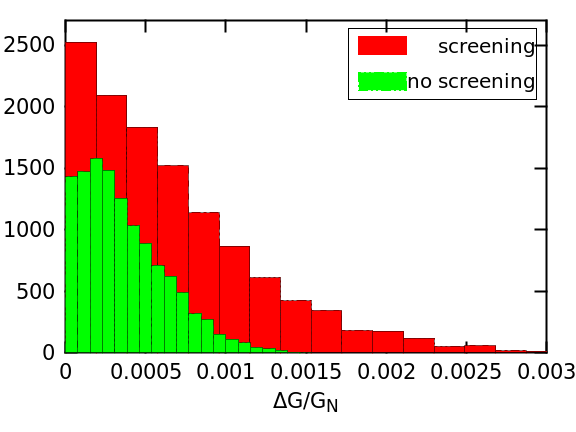}
    \label{fig:6b}
  }
  \subfigure[$\lambda_C = 50$ Mpc]
  {
    \includegraphics[width=0.3\textwidth]{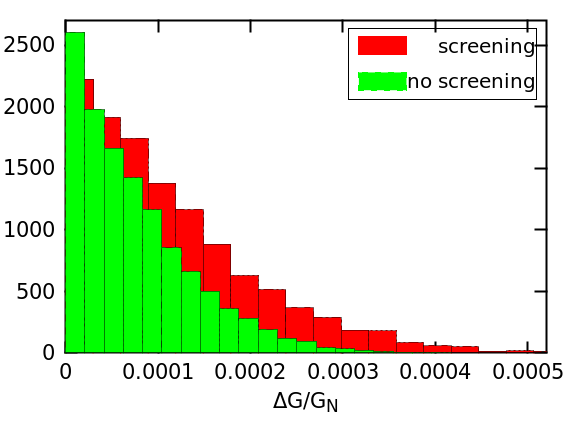}
    \label{fig:6c}
  }
  \caption{Posteriors on $\Delta G/G_N$ for three values of $\lambda_C$ under the fiducial noise model, with and without screening. In Fig.~\ref{fig:6a}, the posterior for the model without screening is at very small $\Delta G/G_N$. For $\lambda_C \lesssim 3$ Mpc a non-zero $\Delta G$ is statistically preferred, but only when screening is included.}
  \label{fig:6}
\end{figure*}

\begin{figure*}
  \subfigure[]
  {
    \includegraphics[width=0.48\textwidth]{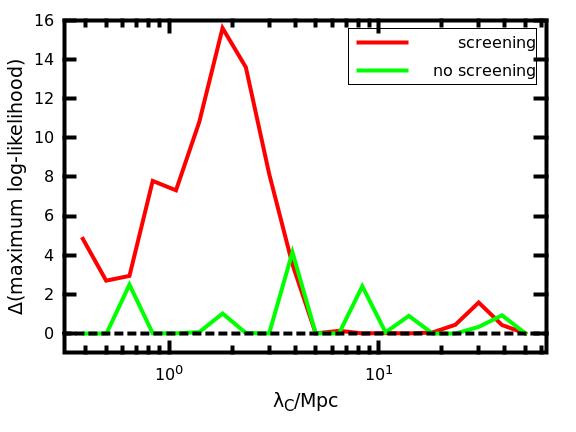}
    \label{fig:like_like}
  }
  \subfigure[]
  {
    \includegraphics[width=0.48\textwidth]{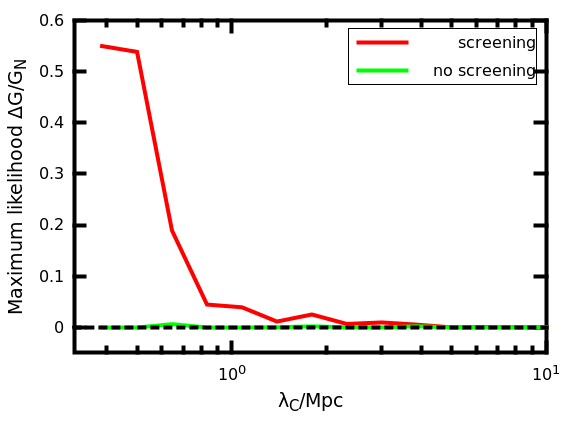}
    \label{fig:like_dg}
  }
  \caption{\textit{Left:} maximum increase in log-likelihood over the case $\Delta G=0$, for any $\Delta G/G_N$, as a function of $\lambda_C$ in the range $0.4-50$ Mpc. We use the fiducial uncertainties and show the results both with (red) and without (green) screening. The model without screening achieves maximum log-likelihood values at most a few larger than GR, which we believe to indicate the noise level of our inference. The model with screening however achieves a maximum increase of $16$ for $\lambda_C = 1.8$ Mpc, with a coherent trend over similar models differing only minorly in fifth-force range. \textit{Right}: best-fit $\Delta G/G_N$ value corresponding to the maximum-likelihood point at each $\lambda_C$. Larger $\Delta G/G_N$ values are inferred at smaller $\lambda_C$ where the radius is lower within which mass contributes to the fifth force, and hence $\vec{a}_5$ and the predicted $r_*$ are smaller.}  
  \label{fig:like}
\end{figure*}

For our first main analysis we switch screening off, fix $\lambda_C$ at one of its 20 values between 400 kpc and 50 Mpc and constrain $\Delta G$ alone. We find the $\Delta G/G_N$ posterior to be peaked at or near $0$ for all $\lambda_C$, indicating no significant deviation from GR. We show in green the $1\sigma$ upper limit on $\Delta G/G_N$ as a function of $\lambda_C$ in Fig.~\ref{fig:constraint}, three example posteriors in Fig.~\ref{fig:6}, and the maximum increase in log-likelihood over the case $\Delta G=0$ in Fig.~\ref{fig:like}. As may be expected, the constraints are tighter for larger $\lambda_C$: $\Delta G/G_N \lesssim 10^{-4}$ $(1\sigma)$ for $\lambda_C=400$ kpc, and $\lesssim$ few$\times10^{-3}$ for $\lambda_C=50$ Mpc. This is because there are more source masses within a larger $\lambda_C$, leading to an enhanced $a_5$ and hence predicted $r_*$, and therefore requiring $\Delta G/G_N$ to be smaller for consistency with the fixed observations. Assuming there is indeed no preference for $\Delta G>0$ in this model, the bumps in the green lines of Figs.~\ref{fig:constraint} and~\ref{fig:like} indicate the noise level of our experiment.

A limit of particular interest for the case without screening is $\lambda_C \rightarrow \infty$, which corresponds to a light scalar field as found for example in Brans--Dicke theory. Extrapolating our constraints to high $\lambda_C$, our analysis may be expected to require $\Delta G/G_N \lesssim 10^{-4}$ in this case, which corresponds to $\omega \gtrsim 500$ in Brans--Dicke. Although not competitive with Solar System results \cite{Alsing}, this constraint derives from a very different gravitational regime. Were $f(R)$ gravity (which requires $\Delta G/G_N = 1/3$) not to employ screening, it would be ruled out to $1\sigma$ across our parameter space, requiring $f_{R0} < 10^{-8}$.

It is important to remember, however, that the physical interpretation of these no-screening results is complicated by the fact that the signal itself relies on a difference between the fifth-force interactions of stars and gas, which most naturally arises when the former are screened and the latter not. Absent screening entirely, the constraints here would require the scalar field simply to couple to the gas and dark matter but not the stars. Our analysis can however be readily extended to the case of arbitrary differential coupling to galaxy mass components. In that case $\vec{a}_5$ in Eq.~\ref{eq:rstar} must be multiplied by the difference in the relative coupling coefficient of the stars and gas, and the constraints on $\Delta G/G_N$ are simply weakened by that factor.

\subsubsection{1D inference with screening}
\label{sec:screening}

With screening switched on, $\lambda_C$ values in the range $\sim0.4-4$ Mpc prefer $\Delta G>0$. We show in red in Fig.~\ref{fig:6} a range of $\Delta G/G_N$ posteriors, and in Figs.~\ref{fig:like_like} and~\ref{fig:like_dg} the maximum log-likelihood increase over GR and best-fit $\Delta G/G_N$ values respectively as a function of $\lambda_C$. The preference for a screened fifth force reflects a typically positive correlation between $\vec{a}_5$ and $\vec{r}_{*,\text{obs}}$. In particular, the data would seem to favour $\Delta G/G_N \sim \mathcal{O}(10^{-2}-10^{-1})$ at $\lambda_C \sim 1$ Mpc ($f_{R0}, \Phi_c \approx 10^{-7}$), with the maximum-likelihood value $\lambda_C = 1.8$ Mpc, $\Delta G/G_N = 0.025$ (also shown in Fig.~\ref{fig:dG3}). This corresponds to $|\Phi_c| = 4.7 \times 10^{-7}$ by Eq.~\ref{eq:f(R)3}, and is not ruled out by any previous analysis. If such a force were realised in nature, neighbouring $\lambda_C$ values would also be expected to be preferred over GR simply because for a given galaxy $\vec{a}_5$ is a continuous function of $\lambda_C$, and hence the predicted $r_*$ values over nearby $\lambda_C$s are correlated. This is the qualitative trend exhibited in Fig.~\ref{fig:like_like}; in Sec.~\ref{sec:valid} we show this trend to be quantitatively as expected also. That no model without screening achieves a meaningful increase in likelihood over GR indicates both that our noise model is robust and that the phenomenon we are observing depends crucially on the environment- and internal structure-dependence of screening. This seems hard to mock up by other means. We test our possible detection in detail in following sections.

The results for the conservative measurement uncertainties (twice fiducial) are shown as the red curve in Fig.~\ref{fig:constraint}. That the $\Delta G/G_N$ constraint is degraded when screening is introduced reflects the fact that when source objects are screened they contribute neither to $a_5$ nor therefore to $r_*$, while when test objects are screened they automatically receive $r_*=0$. The weaker the predicted signal at fixed $\Delta G$, the larger $\Delta G$ may be before it becomes statistically discrepant with the observations. The effect of screening is greater when $|\Phi_c|$ is lower, which by Eq.~\ref{eq:f(R)3} is at lower $\lambda_C$. Conversely, for $\lambda_C \gtrsim 10$ Mpc the majority of source and test points are unscreened in any case, so removing screening has little effect. Note that for $\Delta G/G_N = 1/3$, as in $f(R)$, our conservative error choice requires $\lambda_C \lesssim 500$ kpc, which corresponds to $f_{R0} \lesssim 2 \times 10^{-8}$.

\subsubsection{2D inference with screening}
\label{sec:2D}

\begin{figure*}
  \subfigure[Fiducial]
  {
    \includegraphics[width=0.315\textwidth]{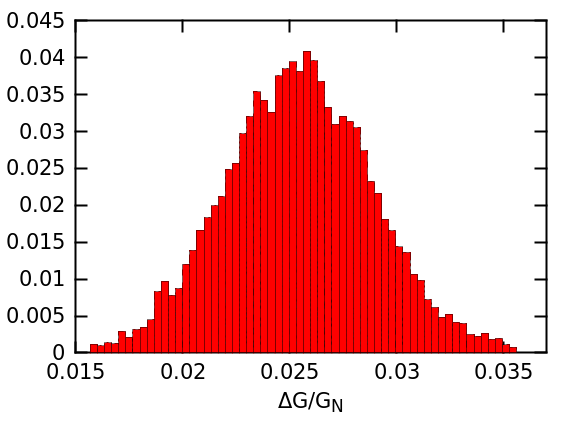}
    \label{fig:2D_dG}
  }
  \subfigure[Fiducial]
  {
    \includegraphics[width=0.315\textwidth]{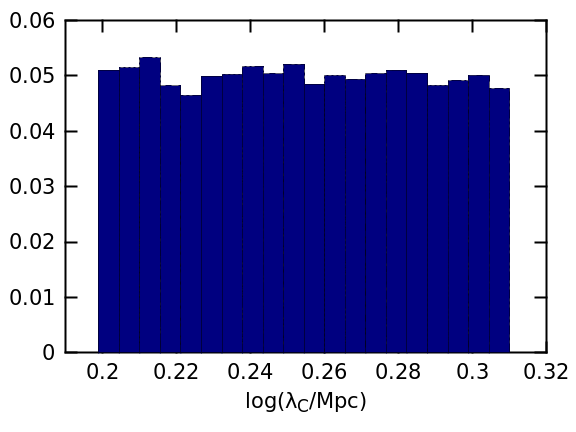}
    \label{fig:2D_lambda}
  }
  \subfigure[Fiducial]
  {
    \includegraphics[width=0.315\textwidth]{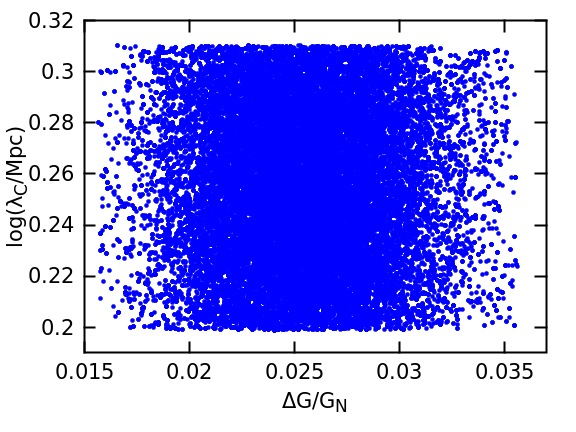}
    \label{fig:2D_comb}
  }
  \subfigure[Conservative]
  {
    \includegraphics[width=0.315\textwidth]{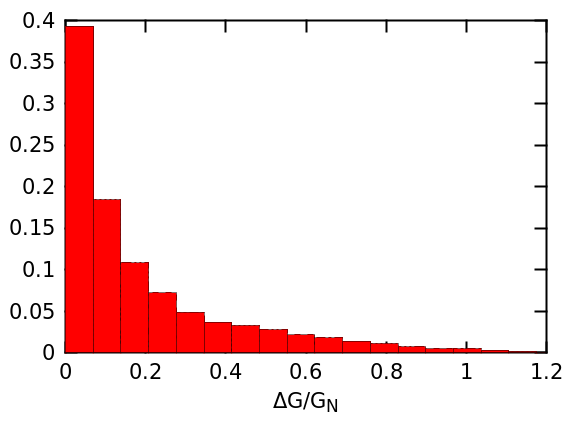}
    \label{fig:2D_dG_2SNR}
  }
  \subfigure[Conservative]
  {
    \includegraphics[width=0.315\textwidth]{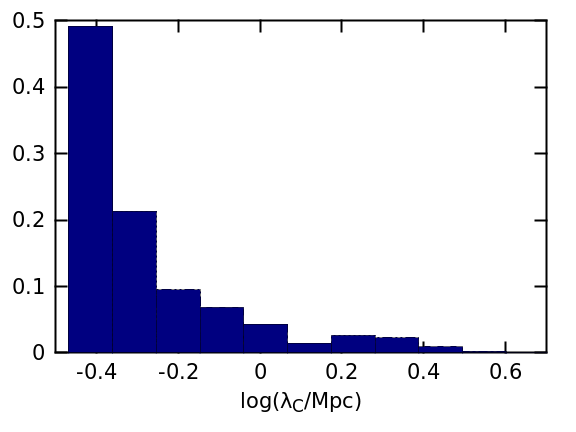}
    \label{fig:2D_lambda_2SNR}
  }
  \subfigure[Conservative]
  {
    \includegraphics[width=0.315\textwidth]{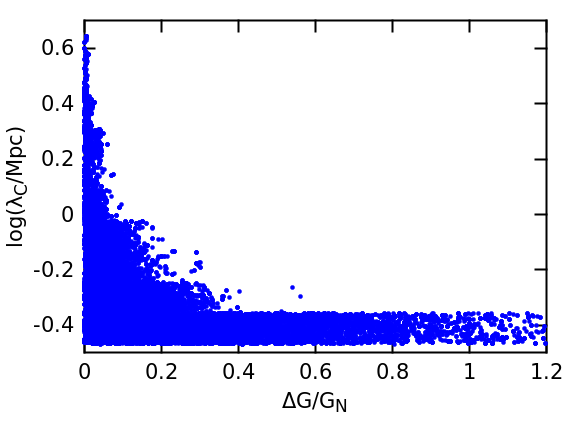}
    \label{fig:2D_comb_2SNR}
  }  
  \caption{Posteriors of $\Delta G/G_N$ and $\lambda_C$ for the full 2D inference, using either the fiducial (top) or conservative (bottom) noise model. The left and centre panels show the marginalised constraints while the right panels show the full 2D posteriors. With the fiducial uncertainties the MCMC picks out the maximum-likelihood point of Fig.~\ref{fig:like_like}, $\lambda_C = 1.8$ Mpc: as this corresponds to a single bin the posterior of $\lambda_C$ is flat. With the conservative uncertainties the preference for low $\lambda_C$ is purely a volume effect arising from the uniform prior, since lower $\lambda_C$ permits a larger range of $\Delta G/G_N$.}
  \label{fig:2D}
\end{figure*}

Finally, we perform the full 2D inference for both $\Delta G$ and $\lambda_C$ in the presence of screening. Under the fiducial noise model the posterior simply picks out the maximum-likelihood solution $\lambda_C = 1.8$ Mpc, with corresponding $\Delta G/G_N = 0.025 \pm 0.003$ (Figs.~\ref{fig:2D_dG}--\ref{fig:2D_comb}). Under the conservative uncertainties (Figs.~\ref{fig:2D_dG_2SNR}--\ref{fig:2D_comb_2SNR}) the preference for low $\lambda_C$ is simply a volume effect: by allowing $\Delta G/G_N$ to span a wider range, a smaller $\lambda_C$ boosts the contribution from the prior. Even with these artificially inflated uncertainties, however, we are able to place the strong constraint $\Delta G/G_N < 0.95$ ($3\sigma$) for any $\lambda_C$ in the range $0.4-50$ Mpc.

\subsection{Validation}
\label{sec:valid}

Fig.~\ref{fig:like_like} appears to show a clear preference for $\Delta G>0$ for $\lambda_C \simeq 1-4$ Mpc in the case where screening is included. Here we document the tests we have performed to validate this detection, including the use of mock data with a $\Delta G/G_N$ value injected by hand, jackknife and bootstrap resampling, and rotation of the measured displacement on the plane of the sky.

\subsubsection{Mock data with $\Delta G/G_N=0$}

\begin{figure}
  \centering
  \includegraphics[width=0.5\textwidth]{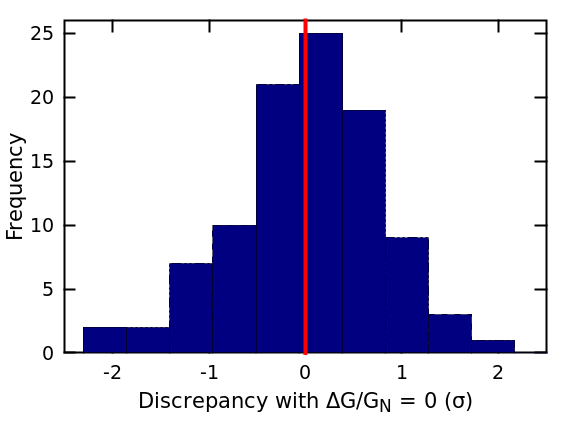}
  \caption{Distribution of deviations of best-fit values of $\Delta G/G_N$ from $0$ (median of $\Delta G/G_N$ posterior divided by standard deviation) derived by fitting a model with $\lambda_C = 1.8$ Mpc to mock data sets with $r_{*,\text{obs},\alpha/\delta}$ randomly scattered around 0 by the uncertainties. This forms a roughly standard normal distribution, as expected for a healthy noise model. By contrast, the deviation for the real data at this $\lambda_C$ is $6.6\:\sigma$, indicating that it is very unlikely to have been generated by a model with noise only.}
  \label{fig:sig_median}
\end{figure}

We begin by generating $100$ mock datasets with $\Delta G=0$, so that $r_{*,\text{obs},i,\alpha/\delta} \sim \mathcal{N}(0,\theta_{i,\alpha/\delta})$. By giving the offset a random direction on the plane of the sky we ensure its complete non-correlation with $\vec{a}_5$ over the sample. We repeat our inference for each mock dataset with the maximum-likelihood model $\lambda_C=1.8$ Mpc, relaxing the prior $\Delta G \ge 0$ and refitting $\theta_{i,\alpha/\delta}$ from the spread in the $r_{*,\alpha/\delta}$ values in bins of $\log_{10}(s)$ (Sec.~\ref{sec:uncertainties}). In each case we calculate the magnitude of the deviation (in $\sigma$) of the best-fit $\Delta G/G_N$ value from 0 by dividing the median of the posterior by the standard deviation. The histogram over all mock data sets is shown in Fig.~\ref{fig:sig_median}. As expected this forms roughly a standard normal distribution, confirming that the offset from $\Delta G=0$ inferred from the real data -- $6.6\:\sigma$ -- would be highly unlikely in the frequentist sense to arise from the noise alone.

\subsubsection{Mock data with maximum-likelihood $\Delta G/G_N$}

\begin{figure*}
  \subfigure[]
  {
    \includegraphics[width=0.315\textwidth]{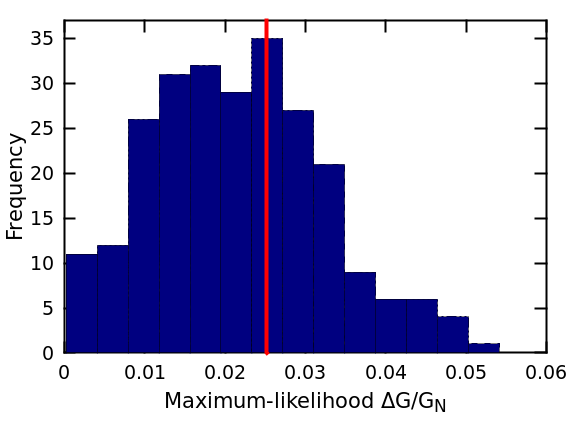}
    \label{fig:mock_dG}
  }
  \subfigure[]
  {
    \includegraphics[width=0.315\textwidth]{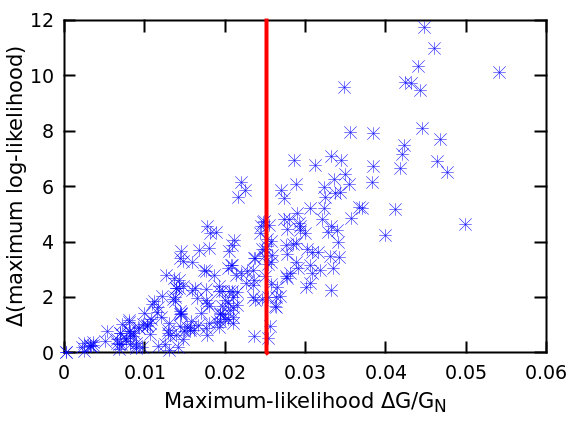}
    \label{fig:mock_dG-dML}
  }  
  \subfigure[]
  {
    \includegraphics[width=0.315\textwidth]{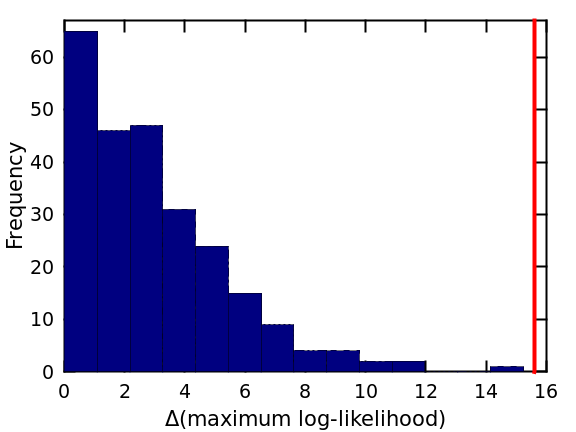}
    \label{fig:mock_dML}
  }  
  \caption{Results of refitting a screened model with $\lambda_C = 1.8$ Mpc to $250$ mock data generated by that model, differing only in noise under the fiducial noise model. Fig.~\ref{fig:mock_dG} compares the distribution of best-fit $\Delta G/G_N$ values to the input value shown by the red line ($0.025$), demonstrating negligible bias. Fig.~\ref{fig:mock_dG-dML} shows that the increase in log-likelihood over the case $\Delta G=0$ is strongly correlated with the $\Delta G/G_N$ value inferred. This increase is however considerably smaller than that achieved in the real data (red line in Fig.~\ref{fig:mock_dML}), indicating that the model is not fully sufficient to account for the observations.}
  \label{fig:mocks}
\end{figure*}

Next we generate mock data with a fifth-force signal injected by hand. This provides an independent test of the validity of the fifth-force interpretation by checking for biases in the reconstructed value of $\Delta G/G_N$. We generate $250$ mock datasets at the maximum-likelihood point of our analysis -- $\lambda_C = 1.8 \: \text{Mpc}, \Delta G/G_N = 0.025$ -- by scattering the expected $r_*$ values\footnote{As in Sec.~\ref{sec:correlations}, we take this to be the mode of the 1000 Monte Carlo model realisations in the case that the galaxy is likely unscreened ($f>0.5$), and 0 otherwise. We have checked that using the mean instead does not lead to appreciably different results.} by the fiducial Gaussian noise in bins of $s$. We refit each one with the $\lambda_C = 1.8$ Mpc model (including recalculation of $\theta_{i,\alpha/\delta}$, as above), and calculate both the maximum-likelihood $\Delta G/G_N$ and the increase in log-likelihood $\Delta\log(\mathcal{L})$ that value achieves over the GR case $\Delta G=0$. The results are shown in Fig.~\ref{fig:mocks}. We find the reconstructed $\Delta G/G_N$ values to be centred around the input value (vertical red line), indicating that our likelihood function picks out a known truth without bias. (We have checked that this holds also for mock data generated by models of different $\lambda_C$.) The considerable variation between mock data sets is due to the dominance of the random noise.

We show in Fig.~\ref{fig:mock_dG-dML} that the $\Delta\log(\mathcal{L})$ values are strongly correlated with best-fit $\Delta G/G_N$, in the sense that mock datasets in which a stronger signal is inferred achieve a larger likelihood increase over GR. The recovered $\Delta\log(\mathcal{L})$ values are however a factor $\sim4$ smaller on average than that in the real data (Fig.~\ref{fig:mock_dML}). This indicates that the relative contribution of signal and noise to our mocks does not perfectly reflect that in the observations: the mock samples have a relatively \emph{weaker} signal than \textit{Alfalfa}. This almost certainly derives from an inadequacy in our noise model, most likely its failure to account for a correlation of the non-fifth-force component of $\vec{r}_*$ with anything other than $s$. This will need to be improved in the future when better priors on both the HI centroid measurement uncertainty and dependence of $\vec{r}_*$ on ``galaxy formation'' physics are available. Other factors which may contribute to disagreement in Fig.~\ref{fig:mock_dML} include an inaccurate $\Phi_c-\lambda_C$ relation (Eq.~\ref{eq:f(R)3} strictly holds only for Hu-Sawicki $f(R)$) and hence to a bias in galaxies' screening fractions as a function of fifth-force range, or to inaccuracies in the galaxy and halo properties input to the model. We discuss these issues further in Sec.~\ref{sec:discussion}.

We have checked also that when the inference is repeated using models with a range of $\lambda_C$ values for the same mock data, the greatest $\Delta\log(\mathcal{L})$ is achieved at the input value $\lambda_C = 1.8$ Mpc. The peak in $\Delta\log(\mathcal{L})$ is however narrower with $\lambda_C$ than in the data, so that neighbouring $\lambda_C$ models have a discrepancy in $\Delta\log(\mathcal{L})$ even greater than a factor of $4$.

\subsubsection{Bootstrap and jackknife resampling}

\begin{figure*}
  \subfigure[Bootstrap $\sigma$]
  {
    \includegraphics[width=0.315\textwidth]{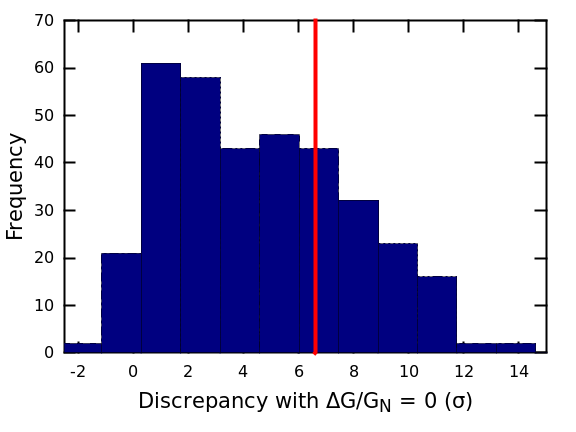}
    \label{fig:boot}
  }
  \subfigure[Jackknife $\sigma$]
  {
    \includegraphics[width=0.315\textwidth]{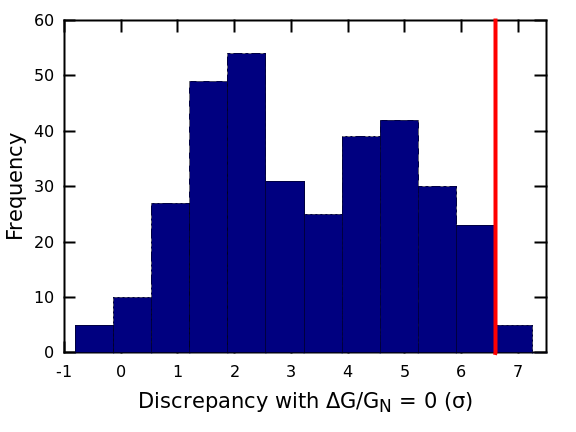}
    \label{fig:jack}
  }
  \subfigure[Maximum-likelihood $\Delta G/G_N$]
  {
    \includegraphics[width=0.315\textwidth]{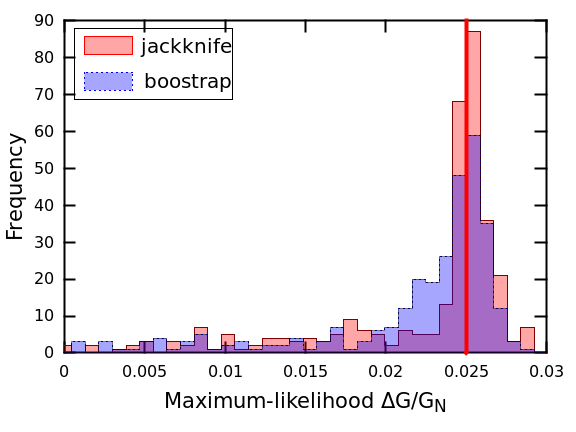}
    \label{fig:jackboot_dG}
  }
  \caption{\textit{Left and centre:} Distributions of deviations as in Fig.~\ref{fig:sig_median} (for the model with $\lambda_C = 1.8$ Mpc), but for datasets derived by bootstrap- or jackknife-resampling the \textit{Alfalfa} sample. Each jackknife dataset contains $70$\% of the full sample, and the vertical red lines show the deviation in the real data ($6.6\sigma$). Although on the whole both sets imply $\Delta G>0$ with lower significance than the full dataset due to reduced statistics, the fact that the majority of resamples achieve a $\gtrsim3\sigma$ detection indicates that the preference for a screened fifth force is not a peculiar property of the \textit{Alfalfa} sample. \textit{Right:} Maximum-likelihood $\Delta G/G_N$ values inferred from the bootstrap and jackknife resamples. These cluster around the \textit{Alfalfa} result, shown as the vertical red line.}
  \label{fig:boot_jack}
\end{figure*}

Next we repeat our analysis with $350$ bootstrap- or jackknife-resampled mock datasets drawn from the $\textit{Alfalfa}$ data. For the jackknife case we retain a random $70\%$ of the galaxies. Again for the model with $\lambda_C=1.8$ Mpc, we calculate the discrepancy in $\sigma$ between the $\Delta G/G_N$ posterior and $\Delta G=0$, and plot the resulting histograms, along with the corresponding maximum-likelihood $\Delta G/G_N$ values, in Fig.~\ref{fig:boot_jack}. The vertical red lines indicate the results in the full \textit{Alfalfa} sample. On the whole the mock datasets have a smaller significance due to the reduced statistics, especially in the jackknife case where the sample size is lower. Nevertheless, the majority of resamples in both cases have a discrepancy with $\Delta G=0$ of $>3\sigma$ and a best-fit $\Delta G/G_N$ value close to $0.025$, indicating that our conclusions do not depend sensitively on peculiar features of the \textit{Alfalfa} sample.

\subsubsection{Rotation of the signal}

\begin{figure*}
  \subfigure[$90\degree$ rotation clockwise]
  {
    \includegraphics[width=0.315\textwidth]{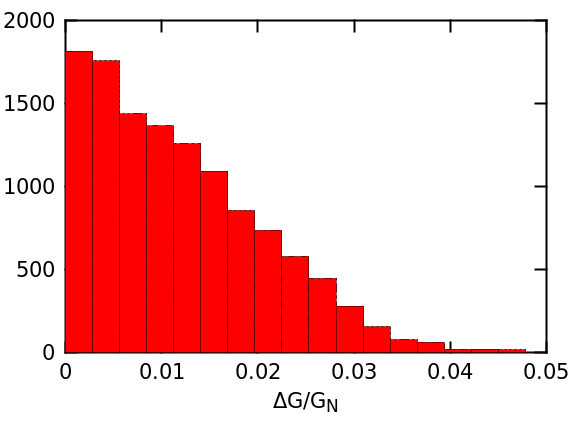}
    \label{fig:90}
  }
  \subfigure[$90\degree$ rotation anticlockwise]
  {
    \includegraphics[width=0.315\textwidth]{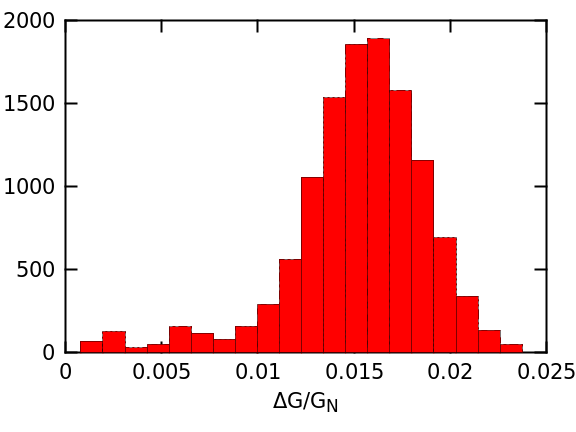}
    \label{fig:90_2}
  }  
  \subfigure[$180\degree$ rotation]
  {
    \includegraphics[width=0.315\textwidth]{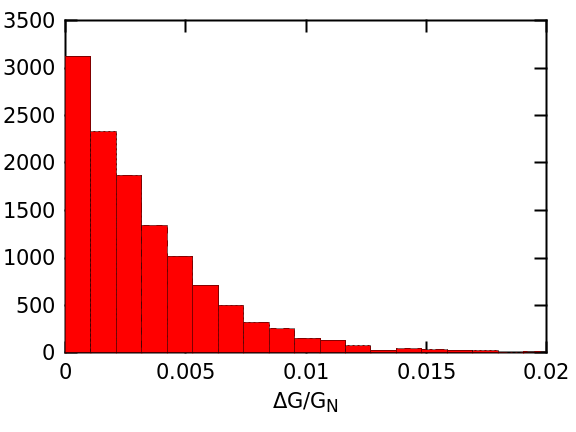}
    \label{fig:180}
  }  
  \caption{$\Delta G/G_N$ posteriors at $\lambda_C = 1.8$ Mpc for the case in which the predicted $\vec{r}_*$ of each galaxy is turned through $90\degree$ in either direction on the plane of the sky (Figs.~\ref{fig:90} and~\ref{fig:90_2}), or $180\degree$ (Fig.~\ref{fig:180}). In the first and third cases the preference for $\Delta G/G_N>0$ is eliminated, indicating no or anti-alignment of the measured $\vec{r}_*$ with the rotated $\vec{a}_5$. In the second a residual correlation remains, although weaker than the unrotated case as evidenced by the lower maximum-likelihood $\Delta G/G_N$ and tail towards small values relative to Fig.~\ref{fig:dG3}. This demonstrates that the observed $\vec{r}_*$ points on the whole in the direction of $\vec{a}_5$.}
  \label{fig:rotation}
\end{figure*}

Finally, we repeat the analysis with the predicted HI-OC displacements (i.e. direction of $\vec{a}_5$) rotated through $90\degree$ or $180\degree$ on the plane of the sky. This is to check that the directions of the observed and predicted signal are indeed positively correlated, as must be the case if part of the signal has a fifth-force origin. The inference of $\Delta G>0$ should be weakened when one of the vectors is turned through $90\degree$, and eliminated when turned through $180\degree$. Fig.~\ref{fig:rotation} shows that this is indeed the case (c.f. the unrotated result in Fig.~\ref{fig:dG3}). We have checked that similar results hold for different $\lambda_C$.

\vspace{5mm}

\noindent In a companion piece we present a similar detection from a largely independent dataset by analysing warps in galactic disks~\citep{Desmond_warp}.

\subsection{Forecasting constraints from future surveys}
\label{sec:forecast}

We now investigate the potential of our constraints for further improvement in the future, focusing solely on the conservative error choice which yields only only upper bounds on $\Delta G/G_N$. To begin, we recall from Sec.~\ref{sec:correlations} (Fig.~\ref{fig:hists}) that the uncertainties in the measured $\vec{r}_*$ are larger by over two orders of magnitude on average than the theoretical uncertainties due to the potential and acceleration fields and test galaxy structure. This implies that the best way to strengthen the constraints is to improve the HI survey; only once the HI uncertainties have been greatly reduced will significant further improvement come from increasing the precision of $\Phi$, $\vec{a}_5$ and $M(<r_*)$.

Our constraint on fifth-force strength $\Delta G/G_N$ therefore depends principally on two parameters, the angular uncertainties $\theta_i$ in the gas--star offsets and the number $N_\text{gal}$ of galaxies observed. While the latter is a property purely of the galaxy survey, the former encapsulates not only the measurement uncertainty in the HI centroid position but also the contribution to the likelihood from non-fifth-force physics. The relative importance of these is not currently known: it may be determined either by increasing the HI resolution -- if the HI-OC offset scales in the same way it derives predominantly from measurement uncertainty -- or by estimating the magnitude of offset that various phenomena in $\Lambda$CDM may be expected to induce. We intend to pursue the second avenue in future work by means of hydrodynamical simulations. In this section, however, we adopt the most optimistic assumption that the entire HI-OC offset is due to measurement uncertainty and may therefore be arbitrarily reduced by future surveys. In practice constraints will be weaker than this due to underestimation of the scatter in HI-OC offset for given HI spatial resolution.

We calculate the $\Delta G/G_N$ constraint as a function of $\theta_i$ and $N_\text{gal}$ by treating them as variables in the generation of mock data sets with $\Delta G=0$, as described in Sec.~\ref{sec:valid}. To begin, we multiply the conservative $\theta_i$ values by a universal scaling factor $10^{-3} < \Theta < 1$. This implies an average HI uncertainty of $\langle \theta \rangle = 36 \: \Theta$ arcsec. Since we only have locations for existing \textit{Alfalfa} galaxies, we consider the dependence of $\Delta G$ on $N_\text{gal}$ for $N_\text{gal}<10,822 \equiv N_\text{Alf}$, the number of galaxies in the \textit{Alfalfa} dataset. We retain a fraction $10^{-3} < f_N < 1$ of these galaxies, chosen at random, and sample $\Theta$ and $f_N$ at $8$ logarithmically equally spaced intervals; to forecast the constraints achievable for $N_\text{gal}>N_\text{Alf}$ we will extrapolate these results.

\begin{figure}
  \centering
  \includegraphics[width=0.5\textwidth]{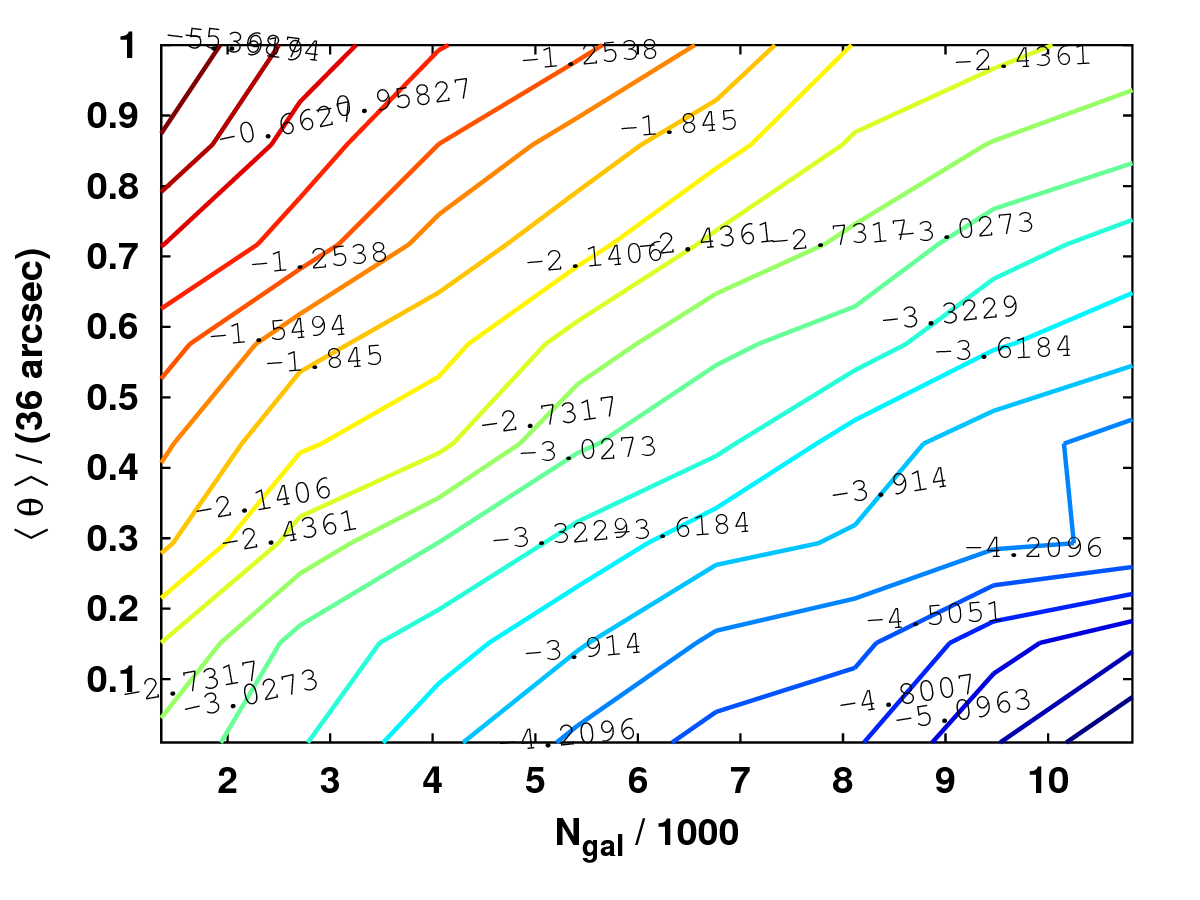}
  \caption{Contour plot forecasting the $1\sigma$ $\log_{10}(\Delta G/G_N)$ constraints obtainable with a dataset of $N_\text{gal}$ galaxies and average angular HI resolution $\langle \theta \rangle = \Theta \times 36$ arcsec, for the screened model with $\lambda_C = 5$ Mpc. We assume the most optimistic case for the noise, in which the entire non-fifth-force part of the signal derives from measurement uncertainty in the position of the HI centroid.}
  \label{fig:contour}
\end{figure}

\begin{figure*}
  \subfigure[]
  {
    \includegraphics[width=0.48\textwidth]{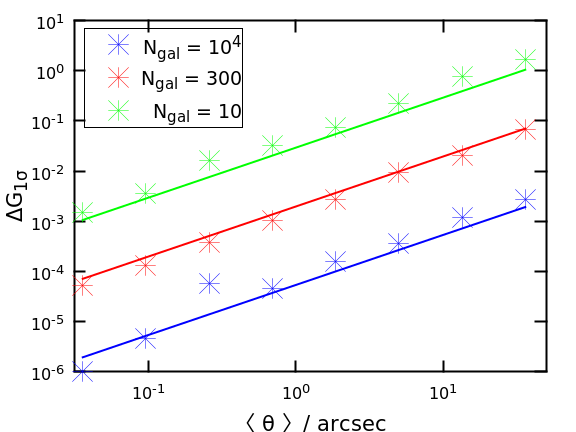}
    \label{fig:arcsec}
  }
  \subfigure[]
  {
    \includegraphics[width=0.48\textwidth]{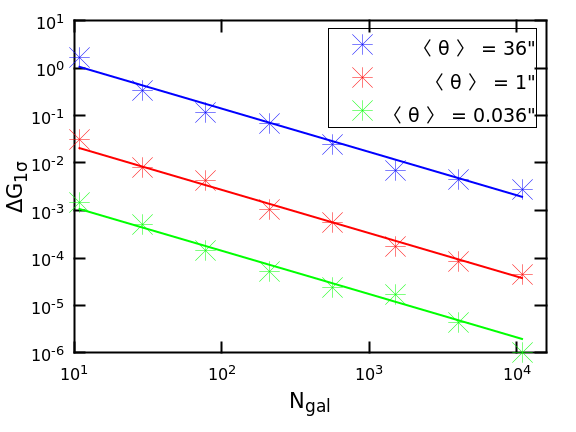}
    \label{fig:Ngal}
  }
  \caption{Variation of $1\sigma$ $\Delta G/G_N$ constraint, $\Delta G_{1\sigma}$, achievable with a dataset of average angular resolution $\langle \theta \rangle$ and size $N_\text{gal}$, for fixed values of the other parameter as indicated in the legend. The solid lines show the fits of Eq.~\ref{eq:fit}.}
  \label{fig:forecast}
\end{figure*}

For a given mock dataset specified by $\Theta$ and $f_N$, we calculate the $1\sigma$ limit on $\Delta G/G_N$ at the fiducial value $\lambda_C=5$ Mpc. To sample both the noise in the mock data and the specific galaxies retained we repeat this procedure 10 times to find an average $\Delta G/G_N$ constraint, $\Delta G_{1\sigma}$. We show the result as a contour plot in Fig.~\ref{fig:contour}, and in Figs.~\ref{fig:arcsec} and~\ref{fig:Ngal} we show the variation of $\Delta G_{1\sigma}$ with $\Theta$ and $N_\text{gal}$ separately (for fixed value of the other parameter) for three example values.

We find $\Delta G_{1\sigma}$ to have approximately power-law dependence on $\langle \theta \rangle$ and $N_\text{gal}$, and therefore fit to it a function of the form
\begin{eqnarray} \label{eq:fit}
\Delta G_{1\sigma} \simeq a \left(\frac{10^3}{N_\text{gal}}\right)^b \left(\frac{\langle \theta \rangle}{1 \ \rm{arcsec}}\right)^c,
\end{eqnarray}
finding best-fit values $\{a, b, c\} = \{8.6 \times 10^{-4}, 0.91, 1.00\}$. We show these fits as the solid lines in Fig.~\ref{fig:forecast}. $\Delta G_{1\sigma}$ therefore falls more rapidly with $N_\text{gal}$ than predicted by the scaling $\Delta G_{1\sigma} \sim 1/\sqrt{N_\text{gal}}$, indicating greater sensitivity to sample size than would apply for the case of $N_\text{gal}$ random draws from a stochastic model under GR. The dependence on $\Theta$ in linear. This fit may be used to forecast constraints on $\Delta G/G_N$ for any future survey: for example (again subject to the proviso that the entire scatter in HI-OC displacement is due to measurement uncertainty), $\{\langle \theta \rangle, N_\text{gal}\} \approx \{10^{-1} \: \text{arcsec}, 10^8\}$, achievable with the `mid' configuration of SKA1 \cite{SKA0, SKA}, should probe $\Delta G/G_N$ to the $10^{-9}$ level. Comparing to figure 4 of~\cite{Adelberger}, we see that at this stage fifth-force constraints by our method will be comparable to lunar laser ranging and planetary probes at much smaller scales; they will also be competitive with those from proposed Solar System tests such as laser ranging to Phobos and optical networks around the sun~\cite{Sakstein_proposal}. Even before SKA, its pathfinders APERTIF~\citep{Apertif} and ASKAP~\citep{Askap} will provide significant further constraining power over \textit{Alfalfa}. We caution however that further modelling work will be necessary to extend our gravitational maps to the distance that this $N_\text{gal}$ requires ($z\sim0.5$), and potentially to model time-variation in parameters such as $f_{R0}$. Given future datasets with more constraining power it will also be desirable to drop or generalise the constraint of Eq.~\ref{eq:f(R)3} in order to treat more general cases of screening.

As we account for the overall statistical impact of sample size but not a systematic shift in gravitational environment or galaxy mass, accuracy of our extrapolated constraints requires the locations and structures of galaxies of lower HI luminosity, observed by future surveys, to be similar on average to those already measured by \textit{Alfalfa}. As fainter galaxies are less likely to self-screen, and may also tend to live in sparser environments, our forecasts are conservative in this regard.

\section{Discussion}
\label{sec:discussion}

\subsection{Noise model}
\label{sec:HI_unc}

The evidence for a screened fifth force depends crucially on the the non-fifth-force part of the likelihood function, modelled by a Gaussian with angular width $\theta_i$. This is poorly known a priori. While average values of $\theta_i$ may be fitted as part of the model, leading to results consistent with our fiducial analysis, correlations with variables other than signal to noise could bias our inference. It is however hard to see how this could eliminate the fifth-force signal we have inferred: this would require that $\theta_i$ correlate not only with galaxies' environments similarly to $\vec{a}_5$, but also with their degree of screening (recall that no model without screening achieves an increase in maximum-likelihood over GR) \emph{and} internal structures in the manner of $\vec{r}_*$ in Eq.~\ref{eq:rstar}. Indeed, our fiducial errors are already conservative in setting the signal to noise ratio of the detection of \emph{any} offset between stars and gas to $\sim1$, a valuable safeguard against erroneously imputing the signal a modified gravity origin in a framework that neglects more mundane phenomena. No additional effect is therefore necessary to explain the observations: the analysis could easily prefer $\Delta G=0$ for all $\lambda_C$. While our conservative error choice is useful for sidestepping the question of $\theta_i$ and hence achieving robust upper bounds on $\Delta G/G_N$, its inadequacy is clear from its severe overprediction of the dispersion of the measured displacements and resultant low overall likelihood.

We have been forced to derive the measurement uncertainties empirically due to insufficient prior information on the precision of the HI centroid location: it is simply too difficult to propagate uncertainties in the telescope pointing, data reduction pipeline and HI-optical cross-correlation, which themselves are poorly known. Future HI surveys, such as those afforded by interferometric observatories such as SKA, may provide better control of these. However, even if information of this nature cannot be used to set informative priors on the HI uncertainties, the increased precision in the centroiding will greatly improve the strength of the test performed here. As demonstrated in Sec.~\ref{sec:forecast}, simply repeating our analysis on an SKA-class sample would probe $\Delta G/G_N$ to $\mathcal{O}(10^{-9})$ for $\lambda_C \sim \mathcal{O}(1-10 \: \text{Mpc})$, easily sufficient to confirm or rule out our putative detection regardless of the precise manner in which uncertainties are modelled.

\subsection{Galaxy formation physics and other systematics}
\label{sec:systematics}

We have been careful to incorporate known uncertainties in the input parameters of our model into probability distributions for those parameters, and hence marginalise over them in the prediction of $\vec{r}_*$. Provided the true parameter values are contained within the priors, these uncertainties cannot cause systematic error in our inference but will simply inflate our posteriors and lead to conservative bounds. Nevertheless, there remain a number of inputs, some of them implicit, whose true values may not lie within our priors. We discuss these here in roughly decreasing order of importance.

Our model assumes that the true $r_{*,\alpha}$ and $r_{*,\delta}$ values (after accounting for noise) arise purely from fifth-force effects. It is however highly likely that these could be non-zero even in the case $\Delta G=0$. This may arise from a number of baryonic processes within a galaxy that affect gas and stars differently, for example hydrodynamical drag, ram pressure and the transfer of energy and momentum by stellar feedback. A realistic model must therefore contain a contingent for $r_* \ne 0$ when $\Delta G=0$ by introducing further parameters describing these baryonic effects. This is present to first order in our model, as the principal component of the likelihood function is simply a Gaussian in $r_{*,\alpha}$ and $r_{*\delta}$ with a width $\theta$ that correlates with the signal to noise ratio $s$ of a galaxy's detection (Sec.~\ref{sec:uncertainties}). Absent modified gravity, larger HI-optical offsets are likely to be caused by baryonic physics, which a larger uncertainty $\theta$ would effectively absorb: in each bin of $s$ the baryonic contribution to the signal adds to the measurement uncertainty in quadrature. The information we glean on modified gravity comes primarily from the relative directions of $\vec{r}_*$ and $\vec{a}_5$, the correlation of their magnitudes across the sample, and the dependence of the screened fraction on the gravitational environment, quantified here by $\Phi$.

Nevertheless, it is important to check that baryonic effects could not generate an environment-dependence of $\vec{r}_*$ similar to the effect of screening. This may be done by applying our inference framework to mock datasets generated by hydrodynamical simulations in which these effects are included. An informative test however requires better than $\mathcal{O}$(kpc) resolution in the central regions of halos, which rules out the majority of current-generation cosmological simulations~\cite{Horizon, EAGLE_1, Illustris, MB-II}. Conversely, zoom simulations do not have the halo numbers required for statistically significant results. Only a few present simulation sets may therefore be of use (e.g.~\cite{Stinson,FIRE, Chan_FIRE, Apostle}). As these simulations differ in the sub-grid models required at the small scales of interest here, it will ultimately be necessary to check several of them. Although the manner in which hydrodynamics affects the internal and environmental properties of halos is at least qualitatively known (e.g~\cite{Hellwing, EAGLE_2, Chisari}), the specific signal in which we are interested has not yet been investigated.

If such effects in $\Lambda$CDM are not sufficient to account for the correlations we attribute here to a screened fifth force, our constraints on $\Delta G/G_N$ must be further validated by examining the impact of baryonic physics in a universe governed by modified gravity. While such simulations have begun in recent years, and confirm that baryonic and gravitational effects on large scale structure and galaxies' internal dynamics are non-linearly coupled and hence not fully orthogonal (e.g.~\cite{Puchwein, Mead}), they focus on fifth-force strengths at least an order of magnitude larger than those that constitute our maximum-likelihood models.

We have assumed the model specified by $\{\Delta G, \lambda_C\}$ to produce a cosmology consistent with $\Lambda$CDM, specifically in terms of its prediction for the smooth density field at $z=0$ and the properties of the halo population. While chameleon- or symmetron-screened modified gravity does alter the expansion history of the Universe and growth rate of structure, and hence the halo mass function and internal structure of halos, its effects in the region of parameter space of most interest here ($\Delta G/G_N \lesssim 0.1$, $\lambda_C \lesssim 5$ Mpc) are small~\citep{Shi, Fontanot, Lombriser_2, Lombriser_rev}. They are likely subdominant to the uncertainties we do model in the smooth density field, galaxy--halo connection and properties of the individual halos themselves. Our method should not therefore be thought a means of testing modified gravity in cosmology -- and has little if any relevance to cosmic acceleration -- but rather as simply a means of seeking fifth forces, of gravitational or non-gravitational origin, on the scales of galaxies and their environments. Screened modified gravity also affects stellar luminosities and hence galaxy mass-to-light ratios~\citep{Davis} as well as the relative kinematics of stars and gas~\citep{Vikram_RC}, but again these effects are small for $\lambda_C \sim 1$ Mpc and would not be expected to skew our inference.

As discussed in~\cite{Desmond}, our incorporation of both halos and a non-halo-based mass contribution leads us to overestimate the total density on $\gtrsim 10$ Mpc scales by $\sim 10$\%. This will cause a similar overestimation of $|\Phi|$, biasing the screened fraction of our test points high and the template signal low. This leads to a weakening of the $\Delta G$ constraints, making our limits conservative. While this effect is partly mitigated by a corresponding overestimation of $\vec{a}_5$, we showed in~\cite{Desmond} that the acceleration is more sensitive than the potential to mass on smaller scales around the test point, which is less susceptible to double counting.

Although we marginalise over the possible galaxy--halo connections for fixed AM proxy and scatter, we do not marginalise over the distributions of these parameters themselves or otherwise consider the possibility that this model inadequately describes our source galaxy population. Given that AM parameters are fairly well constrained by clustering studies, however, this extra uncertainty is unlikely to be significant.

\begin{figure}
  \centering
  \includegraphics[width=0.5\textwidth]{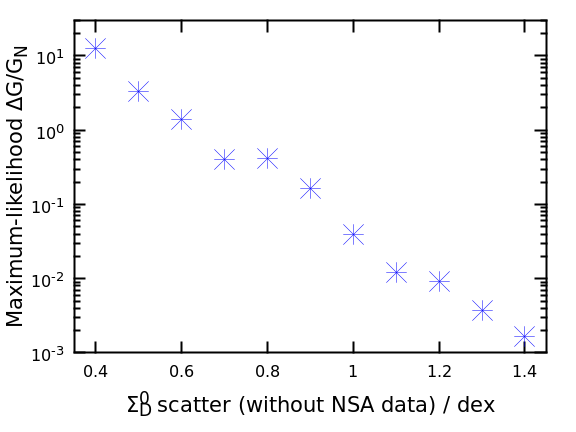}
  \caption{Dependence of maximum-likelihood $\Delta G/G_N$ on the scatter in central dynamical surface density $\Sigma_D^0$ (Eqs.~\ref{eq:CDR}--\ref{eq:rstar2}), in the case where NSA information is not available, for the screened model with $\lambda_C = 1$ Mpc. Our fiducial scatter is 1 dex, as quoted in Table~\ref{tab:params}. The $\Delta G/G_N$ constraint is highly sensitive to this choice, and differs by more than an order of magnitude across the range of a priori reasonable values. This makes our inference of $\Delta G/G_N$ quite uncertain, and the formal uncertainties in Figs.~\ref{fig:dG3} and~\ref{fig:6} potentially misleading. The trend of best-fit $\Delta G/G_N$ with $\lambda_C$ however is not sensitive to this uncertainty because it derives from the scale-dependence of the predicted signal rather than its overall normalisation and scatter.}
  \label{fig:dG_scatt}
\end{figure}

Factors that affect the overall magnitude of the predicted signal (e.g. $\Sigma_D^0$, $h$ and $|\vec{a}_5|$ in Eq.~\ref{eq:rstar2}) are degenerate with the inferred value of $\Delta G/G_N$. As mentioned in Sec.~\ref{sec:more-sampling}, a particularly important quantity is the scatter in central dynamical surface density $\Sigma_D^0$ around the result of Eq.~\ref{eq:CDR}, which affects the magnitude of the predicted $r_*$ and hence our $\Delta G/G_N$ constraints by a relatively large factor. In Fig.~\ref{fig:dG_scatt} we plot the scatter in $\Sigma_D^0$ (for the galaxies without NSA information; the scatter for those with is pinned to half this value) against the maximum-likelihood value of $\Delta G/G_N$ inferred under the fiducial uncertainties for the screened model with $\lambda_C = 1$ Mpc. The slope of the best-fit line is $-3.8$, indicating that a change to the $\Sigma_D^0$ scatter of $0.26$ dex is sufficient to shift the best-fit $\Delta G/G_N$ by an order of magnitude. Were the evidence for $\Delta G>0$ to be confirmed by independent signals or data, it would become important to refine model inputs such as these to better determine $\Delta G/G_N$. Consistency between signals under a generic fifth-force scenario could even be used to constrain them.

By contrast, the $\lambda_C$-dependence of the maximum-likelihood and $\Delta G/G_N$ constraint is degenerate only with factors that scale with $\lambda_C$ in the same way as the predicted $\vec{r}_*$. As this condition is far harder to satisfy due to the peculiar dependence of $\vec{r}_*$ on $\Phi$ and $\vec{a}_5$ (and the further dependences of those on both halo structure and environment), we consider the trend of $\Delta \log(\mathcal{L})$ with $\lambda_C$ to be considerably more robust than the constraint on $\Delta G/G_N$ itself. For example, altering the $\Sigma_D^0$ scatter within the range of Fig.~\ref{fig:dG_scatt} does not significantly affect Fig.~\ref{fig:like_like}.

\subsection{Comparison with the literature}
\label{sec:comparison}

We begin by comparing our work with previous studies at the scales of interest here ($0.4-50$ Mpc; Sec.~\ref{sec:astroscales}) before broadening our discussion to include tests of related models at larger and smaller scales (Sec.~\ref{sec:otherscales}).

\subsubsection{On astrophysical scales}
\label{sec:astroscales}

Much of the groundwork for constraining modified gravity with intra-galaxy data was laid by~\cite{Jain_Vanderplas}, which provided the first calculations of the functional forms and sizes of the expected signals. These authors estimated the magnitude of $\vec{r}_*$ by taking the external field $\vec{a}_5$ to be generated by a single halo towards which the test galaxy is falling. Assuming the test halo has constant density, they argue that $r_*$ could typically be expected to be $\sim1$ kpc for $\Delta G/G_N=1$. This is quantitatively confirmed by Fig.~\ref{fig:1a}. We note however that our estimates of $r_*$ employ significantly more complex and realistic assumptions than those of \cite{Jain_Vanderplas}. First, the fifth-force accelerations for our test galaxies are derived from their full environments rather than a single nearby object (which is itself liable to be screened for lower values of $|\Phi_c|$). Second, we take only the central regions of the test halos to be cored, and calculate their densities from the central baryon surface density rather than by averaging over the entire halo. As halo density presumably falls with $r$ outside the core, this leads us to assign a larger central density than~\cite{Jain_Vanderplas}, reducing the expected $r_*$. However, the significant scatter that we apply to the central densities (1 dex in most cases) causes many halos within each model realisation to have lower density and hence greater predicted $r_*$.

The study most similar to ours is that of~\cite{Vikram} (specifically their section 4), whose authors also searched for evidence of screening in the HI-optical offsets of $\textit{Alfalfa}$ detections. Our work expands on this in three main ways.

\begin{enumerate}

\item{} In~\cite{Vikram}, the fifth-force acceleration and central mass $M(<r)$, and hence the expected $\vec{r}_*$, are not known. This precludes forward-modelling the displacements from theory parameters, and hence constraining them quantitatively. In addition, the direction of $\vec{r}_*$ on the plane of the sky is of no use if the direction of $\vec{a}_5$ is unknown. These authors must therefore rely instead on the average difference in $r_*$ between screened and unscreened subsamples as a summary statistic of the impact of the fifth force. Our likelihood formalism on the other hand uses the full information on a galaxy-by-galaxy basis to quantitatively constrain the theory parameters. In particular, the use of both components of $\vec{r}_*$ on the plane of the sky brings significant additional constraining power.

\item{} Our model for the gravitational field includes not only the mass in halos hosting bright objects, but also the mass in fainter halos and the inter-halo component of the density field~\citep{Desmond}. This eliminates pathological cases in which $\Phi_\text{ex}$ is apparently 0 due simply to incompleteness in the maps of \cite{Cabre} (e.g.~\cite{Vikram_RC}, table 1). It comes further with full point-by-point probability distribution functions for $\Phi$ and $\vec{a}_5$, which enable us to precisely model uncertainties in the predicted signal.

\item{} Our sample size is hugely larger than~\cite{Vikram}'s: 10,822 vs 245 galaxies. This is partly due to the use of a more complete \textit{Alfalfa} catalogue, and partly because our gravitational maps, based on 2M++ rather than SDSS, cover the entire sky. We are not therefore forced to impose a cut on angular position. As shown in Figs.~\ref{fig:contour} and~\ref{fig:Ngal}, a large sample size is critical for beating down the bound on $\Delta G/G_N$.

\end{enumerate}

Reference \cite{Vikram} constrains the average $r_*$ due to modified gravity to be $\lesssim 1$ kpc. Given our results on the average size of the signal in Sec.~\ref{sec:correlations}, this corresponds to $\Delta G/G_N \lesssim 1$ for $\lambda_C = 5$ Mpc. This is significantly weaker than our own constraints due to the extra machinery of point (i) and increased sample size of point (iii), despite the increased screened fractions resulting from the extra mass of point (ii). While our constraints are limited by the observational uncertainty of the HI centroids, those of~\cite{Vikram} were limited primarily by the sophistication of their theoretical modelling. Indeed, our analysis outperforms their forecast that a predicted offset of $0.5$ kpc would be detectable with $24,500$ dwarf galaxies at $\langle \theta \rangle = 24$ arcsec: we are sensitive to a predicted offset of $\sim 0.01$ kpc ($r_* \sim 1$ kpc for $\Delta G/G_N = 1$ and $\Delta G/G_N$ constraints $\sim 0.01$) using $11,000$ galaxies -- with no selection on density -- at $\langle \theta \rangle = 36$ arcsec.

The strongest constraints previously derived on astrophysical scales come from comparing screened and unscreened distance indicators to nearby galaxies. For $|\Phi_c| \lesssim 10^{-6}$ distances derived from screened tip-of-the-red-giant-branch stars would be expected to differ from those of unscreened Cepheids; their consistency implies $\Delta G/G_N \lesssim 0.2$ ($1\sigma$) for $|\Phi_c| > 10^{-6}$ and $\Delta G/G_N \lesssim 1$ for $|\Phi_c| = \text{few} \times 10^{-7}$~\citep{Sakstein}. Similar results are achievable comparing the gas and stellar rotation curves of isolated dwarf galaxies \cite{Vikram_RC}. Despite a weak signal and large observational uncertainties, our inference is several times stronger due to a combination of huge sample size, great range of environments probed, and use of a vector rather than scalar observable which effectively affords two orthogonal signals in the plane of the sky. No previous analysis has reached the level of sensitivity required to probe the region $\{\lambda_C \simeq 1.8 \: \text{Mpc}, \Delta G/G_N \simeq 0.025\}$ in which we find evidence for a signal.

Additional dynamical galactic signals which have not yet been used to place quantitative constraints on fifth forces include warping of stellar disks, asymmetries in disk kinematics and offsets in the rotation curves of stars and gas~\citep{Jain_Vanderplas, Vikram}. Galaxy warps are the subject of~\cite{Desmond_warp}, and the other effects will be studied in future work.

\subsubsection{On smaller and larger scales}
\label{sec:otherscales}

Testing fifth forces on astrophysical scales is a relatively new endeavour: other direct constraints come only from signals at much smaller scales, while cosmological tests are capable of producing independent though indirect and weaker limits. As discussed in Sec.~\ref{sec:intro}, unscreened fifth forces with ranges $10^{-14} < \lambda_C/\text{m} < 10^{-6}$ are strongly constrained by torsion balance experiments, lunar laser ranging and planetary data within the Solar System. Although our $\Delta G/G_N$ constraints are weaker than these by up to 4 orders of magnitude, reflecting the complex nature of astrophysical systems, they provide the possibility of filling in the parameter space of possible distance-dependent (or potential-, acceleration- or curvature-dependent) modifications to GR, which incorporate a fifth force, at the galaxy scale~\citep{Baker}.

The strongest constraint from Solar System tests for $\lambda_C$ as large as that of interest here comes from the Eddington light bending parameter measured by Cassini, which requires $\Delta G/G_N \lesssim 10^{-5}$~\citep{Bertotti}. Our constraint in the absence of screening is 1--2 orders of magnitude worse, and also requires a differential coupling of the fifth force to stars, gas and dark matter. Nevertheless we use a qualitatively different signal, and, as discussed in Sec.~\ref{sec:forecast}, our method is straightforwardly scalable with future data and may be expected to reach the sensitivities of Solar System tests with next-generation surveys.

Without a means of shielding the scalar field from the screening influence of the surrounding dense environment, screened theories cannot be probed within the Milky Way if the screening threshold puts it in the screened regime. This is the case for most viable models. Nevertheless, it is possible to construct laboratory setups, in particular by means of a vacuum chamber with thick walls, in which the chameleon field is decoupled from the field outside~\citep{Burrage}. This allows the creation of regions where the field remains unscreened, enabling tests of the mechanism by means of Casimir forces~\citep{Casimir}, levitated microspheres~\citep{Rider} and atom~\citep{Hamilton} and neutron~\citep{Brax_neutron} interferometry. All of these tests probe short-range fifth forces; see~\cite{Burrage_review, Burrage_review2} for a review.

A number of other tests are possible at cosmological scales (see the review of~\cite{Lombriser_rev}). In general, a cosmological fifth force causes an enhancement of the growth rate of structure at late times, which affects the CMB through the integrated Sachs-Wolfe and Sunyaev-Zel'dovich effects as well as gravitational lensing~\citep{ISW, ISW_2}. Constraints are also possible by means of the galaxy power spectrum~\citep{Dossett}, comparison of dynamical and lensing masses of elliptical galaxies~\citep{Smith}, redshift space distortions~\citep{RSD}, and the abundance~\citep{Schmidt, Ferraro, Cataneo}, internal structure~\citep{Lombriser} and relative dynamical and lensing masses \cite{Terukina, Wilcox} of clusters. These constraints are typically couched in terms of Hu-Sawicki models of $f(R)$, where $\Delta G/G_N = 1/3$ and $f_{R0}$ is related to $\lambda_C$ and $\Phi_c$ through Eq.~\ref{eq:f(R)}. The strongest constraints are at the $f_{R0} \sim 10^{-5}$ level, showing cosmological tests to be significantly less useful for these theories than astrophysical ones.

\subsection{Suggestions for further work}
\label{sec:future}

Our work provides a case study of the use of the gravitational maps of~\cite{Desmond} to constrain aspects of modified gravity on the scale of galaxies and their environments. This scale is characterised by $1 \: \text{kpc} \; \lesssim r \lesssim 50$ Mpc, $10^{-8} \lesssim |\Phi| \lesssim 10^{-4}$, $10^{-17} \lesssim |\vec{a}|/\text{km s}^{-2} \lesssim 10^{-13}$ and curvature $10^{-57} \lesssim K/\text{cm}^{-2} \lesssim 10^{-50}$.

Even continuing to focus solely on fifth-force phenomenology, a range of intra-galaxy statistics besides the separation of stars and gas may be expected to provide new information: as mentioned in Secs.~\ref{sec:intro} and~\ref{sec:astroscales}, a screened fifth force would induce warps in stellar disks and asymmetries in kinematics, as well as boost the rotation velocity of unscreened mass relative to screened~\citep{Jain_Vanderplas}. Provided these signals could be suitably quantified, each could be used to derive constraints on $\Delta G$ and $\lambda_C$ by a method directly analogous to our own. As these analyses would utilise independent data, and possess sensitivity to different types of galaxy in different environments, their results would be complementary. For the evidence for a chameleon- or symmetron-screened fifth force with $\lambda_C \simeq 2$ Mpc presented here to be convincing it must be corroborated (or shown to be below the sensitivity level of current data) by each of these signals. We do this for galaxy warps in \cite{Desmond_warp}. It is also possible to make predictions for future or present data sets from our best-fit $\{\lambda_C, \Delta G/G_N\}$ values, either by taking the maximum-likelihood point or by means of the Bayesian posterior predictive distribution. This would help to avoid confirmation bias in the analysis of that data. As discussed in Sec.~\ref{sec:systematics}, the non-fifth-force part of the likelihood function could be refined by examining the signals in cosmological hydrodynamical $\Lambda$CDM simulations, and the inferences made fully self-consistent by basing the determination of galaxy and large scale structure formation on simulations in modified gravity.

The constraints derivable from all of these signals will straightforwardly strengthen as future galaxy surveys begin observation. Given the amount of information evidently present in this data (as quantified by the strength of our constraints) and the great range of theories testable in tandem through simple phenomenological parametrisations, we propose that gravitational physics be incorporated as a key science driver of these surveys (see also~\cite{Jain_surveys, Jain_Khoury, Jain_NovelProbes}). Constraints based on intra-galaxy statistics may in fact scale better with survey quality than conventional tests based on inter-galaxy statistics due to lower sensitivity to cosmic variance. At the very least they are complementary and come for little extra effort or expense with ongoing survey programmes.

As discussed in~\cite{Desmond}, our methods may find more general use beyond the search for fifth forces. Any theory predicting deviations from GR that correlate with the local gravitational field may in principle be tested; this includes not only the other screening mechanisms listed in Sec.~\ref{sec:intro} (kinetic screening roughly occurs at a critical value of acceleration and Vainshtein at a critical value of curvature;~\cite{Khoury_LesHouches}), but also generic equivalence principle violations (e.g. in Modified Newtonian Dynamics [MOND], whose external field effect imposes an acceleration threshold) and models of dynamical dark energy which entail novel behaviour near the curvature scale of $\Lambda$. Indeed, it is even possible to perform model-independent tests by simply correlating galaxies' gravitational properties with potential modified gravity signals, although the lack of a prediction would preclude forward modelling and Bayesian inference in this case. This method relies on the assumption that any breakdown of GR would impose a different dependence of dynamical behaviour on gravitational environment than would contaminating ``galaxy formation'' effects, as we believe to be the case for the present signal $\vec{r}_*$. To check this assumption a more precise determination of the impact of baryonic physics would be helpful. This programme would build gravity phenomenology from the ground up in galaxies, rather than test specific theories top down.

We made no a priori distinction here between galaxies in different environments, but rather fed the entire \textit{Alfalfa} dataset into our inference framework. This approach is ideal when all galaxies provide useful information on the parameters to be inferred: in our case, even massive galaxies or those in dense environments are of interest because their differences with the remainder inform the constraint on the screening threshold $\Phi_c$. The screened subsample for a given $\Phi_c$ calibrates the behaviour of the signal in the absence of fifth forces. By contrast, information on some modified gravity theories may only be gleaned from galaxies in very specific environments. For example, testing the external field effect in equivalence principle-violating theories would require galaxies in a dominant external field. The prototypical model in this class, MOND, requires $a_\text{ex} > a_0 \approx 1.2 \times 10^{-10} \: \text{m/s}^2$ for the external field to dominate, which is at the extreme high end of the $a_\text{ex}$ distribution of 2M++ galaxies in the local Universe~\citep{Desmond}. In such cases it will be preferable to begin by searching for regions of the local Universe satisfying certain constraints on the gravitational field, and analyse only the subset of galaxies situated therein.

To generalise the discussion in Sec.~\ref{sec:forecast}, we close by listing generic desiderata for surveys and modelling to be useful for probing fundamental physics with galaxies.

\begin{itemize}

\item{} Any test that depends on a galaxy's gravitational environment must make use of a reconstruction of the gravitational field along the lines of~\cite{Desmond}. Improving the precision of the $\Phi$, $\vec{a}$ and $K$ maps requires principally an increase in the limiting magnitude out to which galaxies, the primary mass tracers, are visible. This is particularly important for pushing studies to higher redshift $z \gtrsim 0.05$ where current all-sky surveys become substantially incomplete. This will improve statistics and enhance the range of environments able to be probed. We estimate in~\cite{Desmond} the improvement in the precision of gravitational parameter inferences afforded by upcoming photometric surveys. This information may be augmented by weak lensing data, spectroscopic information and photometry at other wavelengths to further constrain the overall distribution of mass.

\item{} Robust detections of the fifth-force signals described above require low-redshift, high resolution data at multiple wavelengths to distinguish galaxies' different mass components. In general, data quality is more important than sample size, area or redshift range. There is no requirement that data be homogeneous or uniform across the sky, since spatial statistics of the galaxy population as a whole are not in question. This enables smaller surveys and isolated observations to play a larger role than is possible for traditional probes.

\item{} Resolved spectroscopy is necessary to study dynamical signals of new fields, such as differences between the rotation curves of stars and gas, kinematical asymmetries and mass discrepancies. These are complicated in the case of screened theories by the fact that the most common tracer of stellar kinematics, $H\alpha$ emission, derives from the ionised Str\"{o}mgren spheres surrounding stars which are likely to be at least partly unscreened. The screened kinematics of the stars themselves must therefore be derived from stellar absorption lines, which are considerably more difficult to measure with high signal to noise. Spectroscopy will also be useful to tighten constraints on galaxies' central dynamical masses, which provide the restoring forces on stellar centroids that compensate for their insensitivity to $F_5$. The combination $\vec{a}_5 \: \Delta G \: (M(<r)/r^2)^{-1}$ appears in most of the formulae for screening signals \cite{Jain_Vanderplas}.

\item{} When observational uncertainties are small, inference is limited by the precision with which the signals may be theoretically modelled on a galaxy-by-galaxy basis. This determines the width of the fifth-force part of the likelihood function and is set primarily by $M(<r)$ and the gravitational parameters $\vec{a}_5$ and $\Phi$. Our calculation of the latter depends on the BORG inference of the smooth density field, and the precision of that machinery therefore feeds directly into the strength of our results. Improvements are ongoing in the form of the BORG likelihood, method for density field reconstruction and input data used. Our inference of $\vec{a}_5$ and $\Phi$ depends also on the galaxy--halo connection used to assign halos and hence total dynamical mass distributions to galaxies based on their photometry. In the future, additional observables will be brought to bear on constraining halo properties and both the scatter and systematic uncertainty in the galaxy--halo connection will fall.

\end{itemize}

\section{Conclusion}
\label{sec:conc}

We use the observed displacements between galaxies' stellar and gas mass centroids in the $D < 100$ Mpc part of the complete \textit{Alfalfa} catalogue to constrain fifth forces of range 0.4--50 Mpc that couple differentially to stars, gas and dark matter. Our primary case study is screened modified gravity, where the effect of a new scalar field is hidden in high-density environments by means of the chameleon or symmetron mechanism. In this case, stars self-screen, and hence decouple from the fifth force, in otherwise unscreened galaxies. We deploy the gravitational maps of~\cite{Desmond} to identify screened and unscreened regions of the $z<0.05$ Universe for a given fifth-force range $\lambda_C$ and screening threshold $\Phi_c$, and forward-model the spatial offset between the optical and HI emission of each \textit{Alfalfa} galaxy as a function of $\lambda_C$, $\Phi_c$ and fifth-force strength $\Delta G$. We construct a Bayesian likelihood framework to constrain these parameters from the data, propagating fully the uncertainties in the gravitational maps, the galaxy structure inputs to the signal prediction, and the observations.

Models with $\Delta G/G_N \simeq 1$ and $\{\lambda_C, |\Phi_c|\}$ in the range $\{0.4-50 \: \text{Mpc}, \: 2 \times 10^{-8}-4 \times 10^{-4}\}$ produce HI-optical offsets of typical magnitude $\sim$ 0.1--1 kpc for galaxies in the gravitational environments of the local Universe. This is slightly larger when screening is switched off so that all (rather than only unscreened) mass within $\lambda_C$ contributes to the fifth force. We find weak but generally positive correlation between the direction of the observed optical-HI displacement on the plane of the sky and that of an external screened fifth-force field, and between the magnitude of that displacement and the prediction of screened models. We quantify the statistical significance of these correlations under two main assumptions for the noise in the \textit{Alfalfa} data. The fiducial model, which sets the noise Gaussian and equal on average to the scatter in measured optical-HI displacement in bins of signal to noise ratio of the detection, yields $6.6 \: \sigma$ evidence for a component of the offsets deriving from a screened fifth force, with best-fit solution $\lambda_C = 1.8$ Mpc, $\Delta G/G_N = 0.025$. Similar fifth-force models without screening achieve no increase in likelihood over the General Relativistic case $\Delta G=0$. We validate this detection by a number of means, but cannot prove that it may not be caused by other effects such as galaxy formation physics. We will study such possible degeneracies further in future work.

A more conservative (and lower likelihood) model doubles the magnitude of the noise, enabling us to place the robust constraints $\Delta G/G_N \lesssim 0.1$ ($1\sigma$) for $\lambda_C=0.5$ Mpc, and $\Delta G/G_N \lesssim \: \text{few} \: \times 10^{-4}$ for $\lambda_C=50$ Mpc in the case with screening, and $\Delta G/G_N \lesssim \: \text{few} \: \times 10^{-3}$ and $\Delta G/G_N \lesssim \: \text{few} \: \times 10^{-4}$ respectively in the case without. In $f(R)$ gravity (where $\Delta G/G_N = 1/3$) this corresponds to $f_{R0} \lesssim 2 \: \times 10^{-8}$. Even given this artificial inflation of the uncertainties, these constraints are 2 orders of magnitude stronger than those from cosmology, 1 order of magnitude stronger than those from distance indicators and rotation curve tests on galaxy scales, and approaching the strength of Solar System fifth-force tests. This attests to the great constraining power of our data and methodology.

Our study is the first to leverage `big data' from galaxy surveys for gravitational physics by means of an intra-galaxy signal. Such data may be expected to provide crucial information on gravity in the fully non-linear regime in the future, as ever more powerful galaxy surveys are brought to bear. Besides probing qualitatively different regions of gravitational parameter space, intra-galaxy tests should scale in strength more steeply with survey performance than conventional probes due to reduced sensitivity to sample variance and maximal utilisation of information at low redshift where precision is greatest. Under optimistic assumptions for the noise characteristics of the HI data, applying our methodology to radio detections from SKA-class observatories would provide sensitivity to $\Delta G/G_N \sim \mathcal{O}(10^{-9})$ in screened theories with $\lambda_C \simeq 5$ Mpc. Finally, we discuss prospects for a range of additional intra-galactic signals to yield further and complementary information on fundamental physics in this under-explored regime. Crucially, these analyses will confirm or rule out 
the possible screened fifth force detection achieved here.

\section*{Acknowledgements}

This work was supported by St John's College, Oxford. PGF acknowledges support from Leverhulme, STFC, BIPAC and the ERC. GL acknowledges support by the ANR grant number ANR-16-CE23-0002 and from the Labex ILP (reference ANR-10-LABX-63) part of the Idex SUPER (ANR-11-IDEX-0004-02). We thank Martha Haynes for sharing the complete Alfalfa catalogue before public release, Phil Bull for information on SKA and Referee A of an earlier paper for helpful suggestions on validating the detection. Computations were performed at Oxford and SLAC.


\begin{thebibliography}{10}

\bibitem[Adelberger et al.(1991)]{Adelberger_0}
Adelberger E.~G., Heckel B.~R., Stubbs C.~W., Rogers W.~F., 1991, Annual Review of Nuclear and Particle Science, 41, 269

\bibitem[Adelberger et al.(2003)]{Adelberger}
Adelberger E.~G., Heckel B.~R., Nelson A.~E., 2003, Annual Review of Nuclear and Particle Science, 53, 77 

\bibitem[Alsing et al.(2012)]{Alsing}
Alsing J., Berti E., Will C.~M., Zaglauer H., 2012, PRD, 85, 064041 

\bibitem[Babichev et al.(2009)]{kinetic}
Babichev E., Deffayet C., Ziour R., 2009, International Journal of Modern Physics D, 18, 2147 

\bibitem[Baker et al.(2015)]{Baker}
Baker T., Psaltis D., Skordis C., 2015, ApJ, 802, 63 

\bibitem[Begum et al.(2006)]{Begum}
Begum A., Chengalur J.~N., Karachentsev I.~D., Kaisin S.~S., Sharina M.~E., 2006, MNRAS, 365, 1220

\bibitem[\protect\citeauthoryear{Behroozi, Wechsler \& Wu}{Behroozi et al.}{2013}]{Rockstar}
Behroozi P.~S., Wechsler R.~H., Wu H.-Y., 2013, ApJ, 762, 109 

\bibitem[Bernardi et al.(2003a)]{Bernardi}
Bernardi M. et al., 2003, AJ, 125, 1817

\bibitem[Bertotti et al.(2003)]{Bertotti}
Bertotti B., Iess L., Tortora P., 2003, Nature, 425, 374

\bibitem[Blanton \& Roweis(2007)]{kcorrect}
Blanton M.~R., Roweis S., 2007, AJ, 133, 734

\bibitem[Blanton et al.(2011)]{Blanton}
Blanton M.~R., Kazin E., Muna D., Weaver B.~A., Price-Whelan A., 2011, AJ, 142, 31

\bibitem[Brans \& Dicke(1961)]{Brans}
Brans C., Dicke R.~H., 1961, Physical Review, 124, 925 

\bibitem[Brax et al.(2010)]{Brax}
Brax P., van de Bruck C., Davis A.-C., Shaw D., 2010, PRD, 82, 063519

\bibitem[Brax et al.(2012)]{Brax_2}
Brax P., Davis A.-C., Li B., Winther H.~A., 2012, PRD, 86, 044015

\bibitem[Brax et al.(2013)]{Brax_neutron}
Brax P., Pignol G., Roulier D., 2013, PRD, 88, 083004

\bibitem[Burrage et al.(2015)]{Burrage}
Burrage C., Copeland E.~J., Hinds E.~A., 2015, JCAP, 3, 042

\bibitem[Burrage \& Sakstein(2016)]{Burrage_review}
Burrage C., Sakstein J., 2016, JCAP, 11, 045 

\bibitem[Burrage \& Sakstein(2017)]{Burrage_review2}
Burrage C., Sakstein J., 2017, preprint (arXiv:1709.09071) 

\bibitem[Cabr{\'e} et al.(2012)]{Cabre}
Cabr{\'e} A., Vikram V., Zhao G.-B., Jain B., Koyama K., 2012, JCAP, 7, 034 

\bibitem[Carroll et al.(2004)]{Carroll}
Carroll S.~M., Duvvuri V., Trodden M., Turner M.~S., 2004, PRD, 70, 043528

\bibitem[Cataneo et al.(2015)]{Cataneo}
Cataneo M. et al., 2015, PRD, 92, 044009

\bibitem[Chabrier(2003)]{chabrier}
Chabrier G., 2003, PASP, 115, 763 

\bibitem[Chan et al.(2015)]{Chan_FIRE}
Chan T.~K., Kere{\v s} D., O{\~n}orbe J.,Hopkins P.~F., Muratov A. L., Faucher-Gigu{\'e}re C.-A., Quataert E., 2015, MNRAS, 454, 2981 

\bibitem[Chiaverini et al.(2003)]{Stanford}
Chiaverini J., Smullin S.~J., Geraci A.~A., Weld D.~M., Kapitulnik A., 2003, PRL, 90, 151101

\bibitem[Chisari et al.(2018)]{Chisari}
Chisari N.~E. et al., 2018, preprint (arXiv:1801.08559)

\bibitem[Ciotti \& Bertin(1999)]{Ciotti}
Ciotti L., Bertin G., 1999, AAP, 352, 447

\bibitem[Clifton et al.(2012)]{Clifton}
Clifton T., Ferreira P.~G., Padilla A., Skordis C., 2012, PhysRep, 513, 1 

\bibitem[\protect\citeauthoryear{Conroy, Wechsler \& Kravtsov}{Conroy et al.}{2006}]{Conroy}
Conroy C., Wechsler R.~H., Kravtsov A.~V., 2006, ApJ, 647, 201 

\bibitem[Davis et al.(2012)]{Davis}
Davis A.-C., Lim E.~A., Sakstein J., Shaw D.~J., 2012, PRD, 85, 123006 

\bibitem[de Blok(2010)]{cusp-core}
de Blok W.~J.~G., 2010, Adv. Astron., 2010, 789293

\bibitem[DeBoer et al.(2009)]{Askap}
DeBoer D.~R. et al., 2009, IEEE Proceedings, 97, 1507

\bibitem[Decca et al.(2007)]{Casimir}
Decca R.~S., L{\'o}pez D., Fischbach E., Klimchitskaya G.~L., Krause D.~E., Mostepanenko V.~M., 2007, PRD, 75, 077101

\bibitem[Diemer \& Kravtsov(2015)]{Diemer_Kravtsov}
Diemer B., Kravtsov A.~V., 2015, ApJ, 799, 108

\bibitem[De Felice \& Tsujikawa(2010)]{fR2}
De Felice A., Tsujikawa S., 2010, Living Reviews in Relativity, 13, 3

\bibitem[Desmond et al.(2018a)]{Desmond}
Desmond H., Ferreira P.~G., Lavaux G., Jasche J., 2018, MNRAS, 474, 3152 

\bibitem[Desmond et al.(2018b)]{Desmond_PRL}
Desmond H., Ferreira P.~G, Lavaux G., Jasche J., 2018, preprint (arXiv:1802.07206) 

\bibitem[Desmond et al.(2018c)]{Desmond_warp}
Desmond H., Ferreira P.~G, Lavaux G., Jasche J., 2018, preprint (arXiv:1807.11742) 

\bibitem[Dossett et al.(2014)]{Dossett}
Dossett J., Hu B., Parkinson D., 2014, JCAP, 3, 046 

\bibitem[Dutton et al.(2011)]{Dutton11}
Dutton A.~A. et al., 2011, MNRAS, 416, 322

\bibitem[Falck et al.(2015)]{Falck}
Falck B., Koyama K., Zhao G.-B., 2015, JCAP, 7, 049

\bibitem[Famaey \& McGaugh(2012)]{Famaey_McGaugh}
Famaey B., McGaugh S.~S., 2012, Living Rev. Relativ., 15, 10 

\bibitem[Ferraro et al.(2011)]{Ferraro}
Ferraro S., Schmidt F., Hu W., 2011, PRD, 83, 063503 

\bibitem[Fontanot et al.(2013)]{Fontanot}
Fontanot F., Puchwein E., Springel V., Bianchi D., 2013, MNRAS, 436, 2672 

\bibitem[Geha et al.(2006)]{Geha}
Geha M., Blanton M.~R., Masjedi M., West A.~A., 2006, ApJ, 653, 240

\bibitem[Giovanelli et al.(2005)]{Giovanelli_1}
Giovanelli R. et al., 2005, AJ, 130, 2598

\bibitem[Giovanelli et al.(2007)]{Giovanelli_2}
Giovanelli R. et al., 2007, AJ, 133, 2569 

\bibitem[Hamilton et al.(2015)]{Hamilton}
Hamilton P., Jaffe M., Haslinger P., Simmons Q., M\"{u}ller H., Khoury J., 2015, Science, 349, 849

\bibitem[Haynes \& Giovanelli(1984)]{Haynes_84}
Haynes M.~P., Giovanelli R., 1984, AJ, 89, 758 

\bibitem[Haynes et al.(2011)]{Haynes}
Haynes M.~P. et al., 2011, AJ, 142, 170 

\bibitem[Hellwing et al.(2016)]{Hellwing}
Hellwing, W.~A., Schaller, M., Frenk C.~S., Theuns T., Schaye J., Bower R.~G., Crain R.~A., 2016, MNRAS, 461, L11 

\bibitem[Henriques et al.(2015)]{Henriques}
Henriques B.~M.~B., White S.~D.~M., Thomas P.~A., Angulo R., Guo Q., Lemson G., Springel V., Overzier R., 2015, MNRAS, 451, 2663

\bibitem[Hinterbichler \& Khoury(2010)]{Hinterbichler}
Hinterbichler K., Khoury J., 2010, PRL, 104, 231301

\bibitem[Hopkins et al.(2014)]{FIRE}
Hopkins P.~F., Kere{\v s} D., O{\~n}orbe J., Faucher-Gigu{\'e}re C.-A, Quataert E., Murray N., Bullock J.~S., 2014, MNRAS, 445, 581

\bibitem[Hoyle et al.(2001)]{Hoyle}
Hoyle C.~D., Schmidt U., Heckel B.~R., Adelberger E.~G., Gundlach J.~H., Kapner D.~J., Swanson H.~E., 2001, PRL, 86, 1418 

\bibitem[Hu \& Sawicki(2007)]{Hu_Sawicki}
Hu W., Sawicki I., 2007, PRD, 76, 064004 

\bibitem[Hui et al.(2009)]{Hui}
Hui L., Nicolis A., Stubbs C.~W., 2009, PRD, 80, 104002

\bibitem[Jain(2011)]{Jain_surveys}
Jain B., 2011, Philosophical Transactions of the Royal Society of London Series A, 369, 5081 

\bibitem[Jain \& Khoury(2010)]{Jain_Khoury}
Jain B., Khoury J., 2010, Annals of Physics, 325, 1479 

\bibitem[Jain \& VanderPlas(2011)]{Jain_Vanderplas}
Jain B., VanderPlas J., 2011, JCAP, 10, 032 

\bibitem[Jain et al.(2013)]{Jain_NovelProbes}
Jain B. et al., 2013, preprint (arXiv:1309.5389)

\bibitem[Jain et al.(2013)]{Sakstein}
Jain B., Vikram V., Sakstein J., 2013, ApJ, 779, 39

\bibitem[Janz et al.(2016)]{Janz}
Janz J., Cappellari M., Romanowsky A.~J., Ciotti L., Alabi A., Forbes D., 2016, MNRAS, 461, 2367)

\bibitem[Jasche et al.(2010)]{Jasche_10}
Jasche J., Kitaura F.~S., Wandelt B.~D., En{\ss}lin T.~A., 2010, MNRAS, 406, 60 

\bibitem[Jasche \& Wandelt(2012)]{Jasche_Wandelt_12}
Jasche J., Wandelt B.~D., 2012, MNRAS, 425, 1042 

\bibitem[Jasche \& Wandelt(2013)]{Jasche_Wandelt_13}
Jasche J., Wandelt B.~D., 2013, MNRAS, 432, 894 

\bibitem[Jasche et al.(2015)]{Jasche_15}
Jasche J., Leclercq F., Wandelt B.~D., 2015, JCAP, 1, 036 

\bibitem[Jasche \& Lavaux(2018)]{BORG_PM}
Jasche J., Lavaux G., 2018, preprint (arXiv:1806.11117) 

\bibitem[Kaviraj et al.(2017)]{Horizon}
Kaviraj S. et al., 2017, MNRAS, 467, 4739 

\bibitem[Kent et al.(2008)]{Kent}
Kent B.~R. et al., 2008, AJ, 136, 713

\bibitem[Khandai et al.(2015)]{MB-II}
Khandai N., Di Matteo T., Croft R., Wilkins S., Feng Y., Tucker E., DeGraf C., Liu M., 2015, MNRAS 450, 1349 

\bibitem[Khoury \& Weltman(2004)]{chameleon}
Khoury J., Weltman A., 2004, PRD, 69, 044026

\bibitem[Khoury(2010)]{Khoury_review}
Khoury J., 2010, preprint (arXiv:1011.5909)

\bibitem[Khoury(2013)]{Khoury_LesHouches}
Khoury J., 2013, preprint (arXiv:1312.2006) 

\bibitem[Kravtsov et al.(2004)]{Kravtsov}
Kravtsov A.~V., Berlind A.~A., Wechsler R.~H., Klypin A.~A., Gottlober S., Allgood B., Primack J.~R., 2004, ApJ, 609, 35

\bibitem[Lavaux \& Hudson(2011)]{2M++}
Lavaux G., Hudson M.~J., 2011, MNRAS, 416, 2840 

\bibitem[Lavaux \& Jasche(2016)]{Lavaux}
Lavaux G., Jasche J., 2016, MNRAS, 455, 3169 

\bibitem[Lehmann et al.(2015)]{Lehmann}
Lehmann B.~V., Mao Y.-Y., Becker M.~R., Skillman S.~W., Wechsler R.~H., 2015, preprint (arXiv:1510.05651)

\bibitem[Lelli(2014)]{Lelli_1}
Lelli F., 2014, Galaxies, 2, 292 

\bibitem[Lelli et al.(2016)]{Lelli_2}
Lelli F., McGaugh S.~S., Schombert J.~M., Pawlowski M.~S., 2016, ApJL, 827, L19 

\bibitem[Lombriser(2014)]{Lombriser_rev}
Lombriser L., 2014, Annalen der Physik, 526, 259 

\bibitem[Lombriser et al.(2012a)]{Lombriser}
Lombriser L., Schmidt F., Baldauf T., Mandelbaum R., Seljak U., Smith R.~E., 2012, PRD, 85, 102001 

\bibitem[Lombriser et al.(2012b)]{ISW_2}
Lombriser L., Slosar A., Seljak U., Hu W., 2012, PRD, 85, 124038 

\bibitem[Lombriser et al.(2013)]{Lombriser_2}
Lombriser L., Li B., Koyama K., Zhao G.-B., 2013, PRD, 87, 123511

\bibitem[Long et al.(1999)]{Colarado}
Long J.~C., Chan H.~W., Price J.~C., 1999, Nuclear Physics B, 539, 23

\bibitem[Martin et al.(2009)]{Martin}
Martin A.~M., Giovanelli R., Haynes M.~P.,Saintonge, A., Hoffman G.~L., Kent B.~R., Stierwalt S., 2009, AJPS, 183, 214 

\bibitem[McGaugh(2004)]{McGaugh_MDAcc}
McGaugh S.~S., 2004, ApJ, 609, 652 

\bibitem[Mead et al.(2016)]{Mead}
Mead A.~J., Heymans C., Lombriser L., Peacock J.~A., Steele O.~I., Winther H.~A., 2016, MNRAS, 459, 1468

\bibitem[Milgrom(2016)]{Milgrom}
Milgrom M., 2016, PRL, 117, 141101 

\bibitem[Prugniel \& Simien(1997)]{Prugniel}
Prugniel P., Simien F., 1997, AAP, 321, 111

\bibitem[Puchwein et al.(2013)]{Puchwein}
Puchwein E., Baldi M., Springel V., 2013, MNRAS, 436, 348 

\bibitem[Ratra \& Peebles(1988)]{Ratra}
Ratra B., Peebles P.~J.~E., 1988, PRD, 37, 3406 

\bibitem[Reddick et al.(2013)]{Reddick}
Reddick R.~M., Wechsler R.~H., Tinker J.~L., Behroozi P.~S., 2013, ApJ, 771, 30 

\bibitem[Rider et al.(2016)]{Rider}
Rider A.~D., Moore D.~C., Blakemore L.~M., Lu M., Gratta G., 2016, PRL, 117, 101101

\bibitem[Rodrigues et al.(2017)]{Rodrigues}
Rodrigues D.~C., del Popolo A., Marra V., de Oliveira P.~L.~C., 2017, MNRAS, 470, 2410

\bibitem[Saintonge et al.(2008)]{Saintonge}
Saintonge A., Giovanelli R., Haynes M.~P., Hoffman G.~L., Kent B.~R., Martin A.~M., Stierwalt S., Brosch N., 2008, AJ, 135, 588-604

\bibitem[Sakstein(2017)]{Sakstein_proposal}
Sakstein J., 2017, preprint (arXiv:1710.03156)

\bibitem[Samain et al.(1998)]{Samain}
Samain E. et al., 1998, AAPS, 130, 235

\bibitem[Sanders(1990)]{Sanders_MDAcc}
Sanders R.~H., 1990, A\&AR, 2, 1

\bibitem[Santos et al.(2015)]{SKA0}
Santos M., Alonso D., Bull P., Silva M.~B., Yahya S., 2014, Proc. Sci. AASKA14, 21

\bibitem[Sawala et al.(2016)]{Apostle}
Sawala T. et al., 2016, MNRAS, 457, 1931

\bibitem[Schaller et al.(2015)]{EAGLE_2}
Schaller M. et al., 2015, MNRAS, 451, 1247 

\bibitem[Schaye et al.(2015)]{EAGLE_1}
Schaye J. et al., 2015, MNRAS, 446, 521

\bibitem[Schmidt et al.(2009)]{Schmidt}
Schmidt F., Vikhlinin A., Hu W., 2009, PRD, 80, 083505 

\bibitem[S\'{e}rsic(1968)]{Sersic}
S\'{e}rsic J.~L., 1968, Atlas de Galaxias Australes (Observatorio Astronomico, Cordoba, Argentina)

\bibitem[Shi et al.(2015)]{Shi}
Shi D., Li B., Han J., Gao L., Hellwing W.~A., 2015, MNRAS, 452, 3179

\bibitem[Skillman et al.(2014)]{DarkSky}
Skillman S.~W., Warren M.~S., Turk M.~J., Wechsler R.~H., Holz D.~E., Sutter P.~M., 2014, preprint (arXiv:1407.2600)

\bibitem[Smith(2009)]{Smith}
Smith T.~L., 2009, preprint (arXiv:0907.4829) 

\bibitem[Song et al.(2007)]{ISW}
Song Y.-S., Peiris H., Hu W., 2007, PRD, 76, 063517 

\bibitem[Sotiriou \& Faraoni(2010)]{fR}
Sotiriou T.~P., Faraoni V., 2010, Reviews of Modern Physics, 82, 451

\bibitem[Stinson et al.(2013)]{Stinson}
Stinson G.~S., Brook, C., Macci{\`o}, A.~V., Wadsley J., Quinn T.~R., Couchman H.~M.~P., 2013, MNRAS, 428, 129

\bibitem[Terukina et al.(2014)]{Terukina}
Terukina, A., Lombriser, L., Yamamoto, K., Bacon D., Koyama K., Nichol R.~C., 2014, JCAP, 4, 013 

\bibitem[Tully \& Fouque(1985)]{Tully}
Tully R.~B., Fouque P., 1985, ApJS, 58, 67

\bibitem[Vainshtein(1972)]{Vainshtein}
Vainshtein A.~I., 1972, Phys. Lett. B, 39, 393 

\bibitem[Vikram et al.(2013)]{Vikram}
Vikram V., Cabr{\'e} A., Jain B., VanderPlas J.~T., 2013, JCAP, 8, 020 

\bibitem[Vikram et al.(2014)]{Vikram_RC}
Vikram V., Sakstein J., Davis C., Neil A., 2014, preprint (arXiv:1407.6044)

\bibitem[Verheijen et al.(2008)]{Apertif}
Verheijen M.~A.~W., Oosterloo T.~A., van Cappellen W.~A., Bakker L., Ivashina M. V., van der Hulst J. M., 2008, The Evolution of Galaxies Through the Neutral Hydrogen Window, 1035, 265 

\bibitem[Vogelsberger et al.(2014)]{Illustris}
Vogelsberger M. et al., 2014, MNRAS, 444, 1518

\bibitem[Weinberg(1967)]{Weinberg}
Weinberg S., 1967, PRL, 19, 1264 

\bibitem[Wetterich(1988)]{Wetterich}
Wetterich C., 1988, Nuclear Physics B, 302, 668

\bibitem[Wilcox et al.(2015)]{Wilcox}
Wilcox H. et al., 2015, MNRAS, 452, 1171

\bibitem[Williams et al.(1996)]{Williams}
Williams J.~G., Newhall X.~X., Dickey J.~O., 1996, PRD, 53, 6730

\bibitem[Yahya et al.(2015)]{SKA}
Yahya S., Bull P., Santos M.~G.,  Silva M., Maartens R., Okouma P., Bassett B., 2015, MNRAS, 450, 2251 

\bibitem[Yamamoto et al.(2010)]{RSD}
Yamamoto K., Nakamura G., H{\"u}tsi G., Narikawa T., Sato T., 2010, PRD, 81, 103517 

\bibitem[Zhao et al.(2011a)]{Zhao_1}
Zhao G.-B., Li B., Koyama K., 2011, PRL, 107, 071303 

\bibitem[Zhao et al.(2011b)]{Zhao_2}
Zhao G.-B., Li B., Koyama K., 2011, PRD, 83, 044007 

\end{thebibliography}
\end{document}